%% file: main.tex
\documentclass[12pt]{article}
\usepackage{amsfonts}
\usepackage{amssymb}
\usepackage{amsthm}
\usepackage{amsmath}
\usepackage{bbm}
\usepackage{appendix}
\usepackage{array}
\usepackage{booktabs,longtable}
\usepackage[para,online,flushleft]{threeparttable}
\usepackage{threeparttablex}
\usepackage{color,tikz}
\usepackage{color}
\usepackage{dsfont}
\usepackage{enumerate}
\usepackage{epstopdf}
\usepackage{float}
\usepackage{graphicx}
\usepackage{geometry}
\usepackage[hidelinks]{hyperref}
\usepackage{indentfirst}
\usepackage{longtable}
\usepackage{lscape}
\usepackage{multirow}
\usepackage{multicol}
\usepackage{natbib}
\usepackage{rotating}
\usepackage{setspace}
\usepackage{sectsty}
\usepackage{subcaption}
\usepackage{todonotes}
\usepackage{url}
\usepackage{xcolor}
\usepackage{caption}
\usepackage{subcaption}
\usepackage{titlesec}
\usepackage{times}
\usepackage{mathptmx}
\usepackage{chngcntr}
\usepackage{amsmath,amsfonts, amssymb, dsfont, xcolor, comment}
\usepackage{mathtools}
\usepackage{natbib}
\usepackage{epsfig}
\usepackage{setspace}
\usepackage{graphicx, subcaption}
\usepackage{color}
\usepackage{mdframed}
\usepackage{float}
\usepackage[super]{nth}
\usepackage{tabularx}
\usepackage{ragged2e}
\usepackage{makecell}
\usepackage{placeins}
\usepackage{natbib}
\usepackage{etoolbox}  
\usepackage{setspace}
\usepackage{cleveref}
\usepackage{pdflscape}
\usepackage{tabularx}
\usepackage{longtable}
\usepackage{siunitx}
\newcommand{\sym}[1]{{#1}} 

\newtheoremstyle{boldprop}  
  {\topsep}   
  {\topsep}   
  {\itshape}  
  {}          
  {\bfseries} 
  {.}         
  { }         
  {\thmname{#1} \thmnumber{#2}} 
\theoremstyle{boldprop}

\usepackage{bm}

\newcommand{\Xomit}[1]{}

\usepackage[normalem]{ulem}
\usepackage[mathscr]{eucal}

\long\def\/*#1*/{}

\usepackage{chngcntr}

\usepackage{hyperref}
\hypersetup{colorlinks,linkcolor={blue},citecolor={blue},urlcolor={blue}}  
\newcommand{\jc}[1]{{\color{orange} #1 }}

\begin{document}
\newgeometry{top=1in, bottom=1in, left=1in, right=1in} 
\title{\LARGE{{{AI as ``Co-founder'':\\} 
GenAI for Entrepreneurship}}
\thanks{We are grateful to Petra Todd for her insightful comments and suggestions. We thank Yao Li, Xingyu Lu, Fangzhuo Yang, Dibo Peng, and Zhihan Zhou for outstanding research assistance. Zhu acknowledges financial support from the Tsinghua University Initiative Scientific Research Program (No. 2022Z04W02016) and the Tsinghua University School of Economics and Management Research Grant (No. 2022051002). All remaining errors are our own.}
}
\vspace{-0.5in}
\date{December 6, 2025}

\author{ 
\hspace{5mm} Junhui Jeff  Cai \footnote{Mendoza College of Business, University of Notre Dame, jcai2@nd.edu} 
\hspace{5mm} Xian Gu \footnote{Department of Finance, Durham University Business School, xian.gu@durham.ac.uk}  
\hspace{5mm} Liugang Sheng \footnote{Department of Economics, The Chinese University of Hong Kong, lsheng@cuhk.edu.hk}  \\
\hspace{5mm} Mengjia Xia \footnote{Department of Economics, The University of Pennsylvania, xiax@sas.upenn.edu}  
\hspace{5mm} Linda Zhao  \footnote{Wharton School, The University of Pennsylvania, lzhao@wharton.upenn.edu}  
\hspace{5mm} Wu Zhu  \footnote{Department of Finance, SEM, Tsinghua University, zhuwu@sem.tsinghua.edu.cn} }

\maketitle 
\vspace{-0.3in}
\maketitle

\begin{abstract}
\singlespacing  
This paper studies whether, how, and for whom generative artificial intelligence (GenAI) facilitates firm creation.
Our identification strategy exploits the November 2022 release of ChatGPT as a global shock that lowered start-up costs and leverages variations across geo-coded grids with differential pre-existing AI-specific human capital. Using high-resolution and universal data on Chinese firm registrations by the end of 2024, we find that grids with stronger AI-specific human capital experienced a sharp surge in new firm formation---driven entirely by small firms, contributing to 6.0\% of overall national firm entry. Large-firm entry declines, consistent with a shift toward leaner ventures. New firms are smaller in capital, shareholder number, and founding team size, especially among small firms. The effects are strongest among firms with potential AI applications, weaker financing needs, and among first-time entrepreneurs. Overall, our results highlight that GenAI serves as a pro-competitive force by disproportionately boosting small-firm entry.

\medskip


\noindent\textbf{JEL Classification}: L26, O33, L53, G30 \\
\noindent\textbf{Keywords}: Generative AI; Entrepreneurship; Firm Entry; Human Capital; Technological Change; Spatial Economics


\end{abstract}

\restoregeometry

\thispagestyle{empty}
\addtocounter{page}{-1}

\newpage
\setstretch{1.4}

\section{Introduction}

Technological evolution underpins long-run growth, driving entrepreneurship as ``creative destruction'' \citep{schumpeter1943capitalism} that turns new technologies into new firms and markets \citep{aghion2018causal}. 
The recent meteoric rise of \textit{generative artificial intelligence} (GenAI) was unprecedented: OpenAI's ChatGPT (GPT-3.5), released in November 2022, reached roughly 1 million users in five days and about 100 million monthly active users within two months, outpacing any prior platforms.\footnote{TikTok reached 100 million users in around 9 months and Instagram in around 2.5 years. \href{https://finance.yahoo.com/news/chatgpt-on-track-to-surpass-100-million-users-faster-than-tiktok-or-instagram-ubs-214423357.html}{Yahoo Finance}.}
Crucially for entrepreneurship, GenAI differs from earlier waves because it combines broad task generality
with easy accessibility via a universal natural-language interface \citep{eloundou2024gpts}. In practice, it automates language-based information-processing (reasoning, synthesis, coding, and generative), alongside creative and managerial tasks, thereby lowering the barriers to capabilities that were once capital-intensive or concentrated among specialized labor and well-resourced organizations.
GenAI thus can serve as a co-founder and enables entrepreneur to function as a small, multi-skilled team.
As a result,
teams shrink, fixed costs fall, and entry accelerates.
“AI-native” ventures such as \text{Midjourney}\footnote{https://www.midjourney.com/home}, \text{Cursor}\footnote{https://cursor.com}, and \text{Perplexity}\footnote{https://www.perplexity.ai} exemplify this shift.
Against this backdrop, we ask: \emph{does the diffusion of GenAI increase firm creation, through which mechanisms, and who benefits most?}

In this paper, we study these questions by exploiting the November 2022 release of ChatGPT as a sharp shock to the salience and availability of GenAI. Motivated by recent evidence that human capital plays a central role in the diffusion and adoption of AI technologies \citep{babina2023firm, gofman2024artificial, andreadis2025local,zhu2024divergent}, our identification strategy leverages \emph{within-city} variation across geocoded grids with differential pre-existing AI-specific human capital, proxied by the number of AI invention patents filed between 2010 and 2019. To implement this design, we combine two comprehensive datasets: (i) universal firm registration records in China covering more than 12 million newly established firms from 2021–2024, and (ii) the complete set of AI invention patents filed between 2010–2019. We geolocate each new firm to a hexagonal H3 grid cell—at roughly 5~km$^2$ spatial resolution—and aggregate pre-2020 AI patents to the same cells using assignee addresses. This procedure yields a novel, nationwide H3 grid-by-quarter panel comprising 166{,}156 spatial cells, enabling a highly granular comparison of how GenAI reshaped firm creation across grids with varying stocks of AI-relevant human capital.

We focus on China for both scale and relevance: it is a major AI hotbed for entrepreneurship and research, and a major economy with detailed nationwide geo-coded firm registration microdata. 
We identify over 12 million new firm registrations from 2021–2024, including over 5.5 million \emph{small} firms (registered capital $<1$ million RMB) and over 6.5 million \emph{large} firms, and classify 340,771 AI-related patents from 2010 to 2019 following the methodology of \cite{fang2025ai}. Our key finding is a sizable surge in new firm formation following the release of ChatGPT in 2022, concentrated in areas with more AI-specific human capital, and driven primarily by small, resource-constrained, AI downstream firms. Our results suggest that GenAI can function as a ``co-founder,'' providing domain knowledge as well as managerial and operational know-how, which reduces entry barriers and favors smaller, leaner organizational forms.




Our baseline empirical identification relies on a novel \textit{difference-in-differences (DiD)} framework that exploits fine-grained spatial heterogeneity in AI-specific human capital within cities before 2020. Specifically, we define a grid with high AI-specific human capital (high AI-exposure) if this grid has at least one AI-related invention patent filed between 2010 and 2019. Next, we compare changes in firm entry before and after the release of ChatGPT between grid cells with high- and low-AI human capital within the same city. The specification includes both \emph{grid-by-calendar-quarter} and \emph{city-by-quarter} fixed effects. The former absorbs seasonal patterns in firm registrations as well as any time-invariant unobservables specific to each grid, while the latter controls for all \emph{time-varying city-level factors}, including local policy initiatives, business environment shifts, industrial restructuring, and macroeconomic fluctuations. This fixed-effect structure ensures that identification arises solely from \emph{within-city, within-quarter} variation across neighboring grids that differ in pre-existing exposure to AI.

This design provides several advantages for causal inference. First, by exploiting micro-geographic variation, it mitigates concerns that our estimates are driven by cross-city differences in economic development, infrastructure, or administrative quality. Second, because the diffusion of ChatGPT represents a global, near-synchronous technology and information  shock, the temporal variation in exposure is plausibly exogenous to local entrepreneurial environment and activities. Third, our approach isolates the role of \emph{AI-specific human capital}—rather than general innovative capacity—by comparing otherwise similar grids within the same urban environment. Remaining identification concerns, such as local spillovers, differential pretrend, confoundedness, and robustness are addressed through a series of placebo and robustness tests discussed later in the paper.



We find that the diffusion of GenAI triggered a pronounced surge in new firm creation across China. Grids with stronger pre-existing AI-specific human capital experienced a sharp and persistent rise in firm entry following the release of ChatGPT. The magnitude of this effect is economically meaningful: high-AI grids recorded roughly five additional new firms per grid–quarter relative to low-AI grids.
Aggregated across all high-AI grids, this translates into approximately 51,000 additional new firms per quarter, accounting for about 6.0\% of total nationwide firm entry after ChatGPT’s release. Event-study estimates show no evidence of differential pre-trends between high- and low-AI grids prior to the shock, supporting a causal interpretation of the results.

A defining feature of this entrepreneurial boom is its asymmetry by firm size. The increase in entry is driven entirely by small, resource-constrained firms, while entry by larger firms declines. 
This pattern is consistent with evidence that GenAI substantially lowers fixed organizational and managerial costs and increases efficiency, thereby effectively lowering the minimum viable scale of new ventures and allowing them to operate with less labor and external finance. 

Further evidence shows that the largest increases in firm entry arise in AI-downstream sectors that apply AI in core production or service—such as retail, business services, online platforms and other digital-service  industries—where GenAI tools can be readily applied to a broad set of product-developing, information-processing, customer-facing, and back-office activities. By contrast, the ChatGPT shock slightly reduces firm entry in traditional and capital-intensive industries including construction and manufacturing, indicating that the sectors whose core of business activities rely on knowledge work, creativity, marketing, and digital service are most responsive to GenAI.

We further investigate the underlying mechanisms. First, the share of serial entrepreneurs who have started at least one firm in the previous three years declines in AI-intensive areas after the release of ChatGPT, indicating that GenAI facilitates entry by first-time founders who traditionally lack experience and managerial or operational know-how. Second, new firms are created with fewer shareholders, reflecting reduced financial constraints at entry as founders rely less on external capital or co-investors. Third, founding teams become smaller, consistent with AI substituting for early-stage managerial and specialized labor and enabling much faster prototyping, thereby allowing leaner organizational forms.
Evidence from repeat founders further corroborates this substitution effect in firm scale: serial entrepreneurs deliberately launch smaller ventures in the post-ChatGPT period, suggesting that GenAI not only expands the extensive margin of entry but also shifts the optimal scale of new firm creation towards leaner organizations.

We conduct a series of placebo exercises to verify that the entrepreneurial response to GenAI is driven by \emph{AI-specific human capital} rather than existing local innovation capability or entrepreneurial trends. 
To separate AI-specific from general innovative capabilities, we replace AI-related patenting with non-AI patenting, a proxy for general innovation capabilities. The estimated treatment effects largely vanish: grids with strong non-AI innovation capacity do not experience comparable post-ChatGPT surges in new firm formation. 

To address the concern that our estimates might simply reflect pre‑existing differences in entrepreneurial activity—rather than AI‑specific human capital—we orthogonalize pre‑2019 firm entry with respect to AI patenting at the grid level. Specifically, we regress pre‑2019 entrepreneurial activity on historical AI patents and construct residuals that purge the variation associated with AI‑specific human capital. We then re‑estimate our design using these residuals as the placebo “treatment.” The resulting effects shrink by roughly 80–85\%, indicating that our baseline findings are not mechanically driven by pre‑existing entrepreneurial vibrancy, but rather by the component of AI exposure that reflects AI‑specific human capital.

As an additional falsification check, we randomly reassign the “high-AI” label across spatial grids and re-estimate our baseline specification for 100 times. The distribution of the estimated coefficients centers tightly around zero, indicating that our baseline results are due to chance spatial correlations or model mis-specification.

Finally, robustness checks confirm that our results are not driven by regional composition, sample design, or measurement choices. Excluding first-tier provinces such as Beijing, Shanghai, and Guangdong yields similar results, indicating that the effects extend well beyond China’s major innovation hubs. Restricting the sample to AI-active grids, matching AI-intensive locations with nearby non-AI counterparts within the same city, and varying the threshold for small firm definition all produce consistent results: grids with stronger pre-existing AI capabilities continue to experience large and statistically significant increases in small-firm entry, alongside corresponding declines in large-firm entry, after the release of ChatGPT. 
We further use a matching estimator that matches AI-grids with non-AI grids within the same city and obtain similar results, suggesting that the effect is not driven by the national or local government industrial policies that support AI as a strategic industry. 

Taken together, our results provide consistent evidence that GenAI relaxes key frictions that traditionally constrain new business formation, including prior entrepreneurial experience, financing, and managerial labor.
These effects are concentrated in regions with stronger AI-specific human capital, where potential founders are in a better position to recognize, adopt, and deploy GenAI, highlighting the importance of human capital and technology-skill complementarities in shaping the real economic implications of general-purpose technologies. 
With GenAI serving as a “co-founder” that allows small teams to replicate capabilities once requiring larger organizations, our findings suggest that GenAI has already begun to reshape entrepreneurial landscape: by democratizing who can start a firm, how firms are organized, and the scale of resources they require to operate.

\paragraph{Contribution to the literature.} Our study contributes to several strands of literature at the intersection of AI, firm dynamics, and labor markets. The first is on firm entry. Firm entry is a vital engine for economic growth, fueling innovation, competition, job creation, and the efficient reallocation of resources. The theoretical importance of this process is rooted in the Schumpeterian concept of ``creative destruction," which posits that long-run economic progress stems not from gradual optimization but from a disruptive process where new firms—armed with innovative products, technologies, or firm models—displace less efficient incumbents \citep{schumpeter1943capitalism}.\footnote{Modern economic theory has formalized this dynamic, with foundational models demonstrating how entrants act as the primary agents of this disruption, driving a selection process that filters out unproductive firms and sustains aggregate growth \citep{jovanovic1982selection, hopenhayn1992entry, aghion1992model}.} A vast body of empirical work, leveraging large-scale firm- and plant-level data, has confirmed that the process of entry and exit is a primary driver of aggregate productivity gains, as resources are reallocated from exiting firms to more productive entrants \citep{foster2001aggregate, brandt2012creative}. Furthermore, it is young firms, not simply small ones, that account for a disproportionate share of net job creation \citep{haltiwanger2013creates}.\footnote{Conversely, a decline in entrepreneurship can signal economic stagnation. For instance, \citet{decker2014role} document a decline in firm dynamism in the US and suggest this trend—characterized by lower startup rates and weaker growth among young firms—is a likely contributor to the nation's recent sluggish productivity growth.} 

A key question in entrepreneurship research is the extent to which technological breakthroughs contribute to firm creation. While they are widely recognized as powerful catalysts, their precise quantitative impact remains difficult to measure. For example, the diffusion of the personal computer fostered the rise of pioneers like Microsoft and Apple, and the internet's commercialization in the mid-1990s triggered a wave of entrepreneurship that produced today's tech giants, including Google, Amazon, and later, Facebook. The primary obstacle to measurement has been methodological: the slow and endogenous diffusion of these past technologies presents significant identification challenges for isolating their causal impact on startup activity. 

The launch of ChatGPT, however, offers a stark contrast and a unique opportunity. Its arrival was largely unexpected, and its adoption has been unprecedentedly rapid, creating a natural quasi-experimental setting. Moreover, since the technology originated in the US, its arrival in China—and the subsequent boom in domestic large language model (LLM) development—represents a plausibly exogenous shock for local entrepreneurs. This context allows researchers to more cleanly identify the causal effects of a major technological breakthrough on firm creation. Our paper addresses this critical gap by providing one of the first pieces of systematic evidence on how technological breakthroughs, such as GenAI, facilitate firm entry, using the universe of newly registered firms in China between 2021 and 2024.

The second body of literature explores the impact of AI on firm dynamics and identifies distinct channels through which it enhances growth and productivity. For instance, \cite{babina2024artificial} find that AI-investing firms in the US experience higher growth in sales, employment, and market valuations, driven primarily by increased product innovation. Complementary evidence from European firm-level data shows that AI innovation and adoption also directly promote firm output and productivity \citep{czarnitzki2023artificial, da2024productivity}. More recently, the boom in GenAI has spurred a new wave of research, with several studies demonstrating that it increases worker productivity across various occupations and promotes sales and productivity of online retail \citep{fedyk2022artificial, brynjolfsson2025generative, kanazawa2025ai, fang2025generative}, while others find that the release of GPT boosted the market valuation of firms with high exposure to the technology \citep{eisfeldt2025generative, bertomeu2025impact}. 

While existing research focuses on incumbent firms and worker-level effects, the impact of AI on market entry remains unexplored. Moreover, the net effect of GenAI on firm creation is theoretically ambiguous. On one hand, GenAI could strengthen market concentration by favoring large, resource-rich incumbents, thereby stifling new entry. On the other hand, it could lower entry barriers by democratizing access to sophisticated tools. Our study addresses this critical question by providing the first large-scale, systematic evidence on how GenAI affects firm entry, drawing on the universe of newly registered firms in China between 2021 and 2024. Contrary to concerns about the increasing market power of large incumbents \citep{babina2024artificial}, we find that GenAI predominantly encourages the entry of small firms. This finding has important implications for market dynamics, suggesting that GenAI may serve as a pro-competitive force that reduces concentration rather than reinforcing the dominance of incumbents.

Third, we also engage with the long-standing debate on the ``race'' between machine and human, and in particular the tension between technological progress enhancing labor productivity and displacing human labor.
While previous waves of automation, such as industrial robotics, primarily impacted low- and middle-skilled blue-collar workers in routine manual tasks \citep{acemoglu2020robots, acemoglu2022artificial}, the advent of GenAI has extended this fear to high-skilled and white-collar professions, though the impact may take time to materialize. By performing sophisticated cognitive and knowledge work, GenAI's impact is likely to be highly concentrated in skilled occupations and high-pay jobs \citep{eloundou2024gpts, hampole2025artificial}. Indeed, emerging evidence indicates that this substitution effect often materializes not as layoffs, but as reduced hiring for entry-level skilled roles, creating significant barriers for new labor-market entrants \citep{lichtinger2025generative}. In contrast to this focus on the substitution effect, our study investigates the role of GenAI in facilitating entrepreneurship in boosting startup and firm creation. We argue that GenAI can reduce the costs associated with starting a firm by automating tasks like coding, marketing content creation, legal document drafting, and firm plan development. This may reduce the initial need for large teams and managerial capital. Thus, GenAI can serve as a powerful engine for entrepreneurship, particularly for inexperienced founders and firms in AI-downstream industries. This entrepreneurial channel is crucial, as new firm creation is a primary engine for net job growth and innovation, offering a vital counterweight to the potential for AI-driven job displacement \citep{haltiwanger2013creates}. Moreover, we also find that geo-coded areas with high AI-specific human capital, not the general capacity, benefit the most from the positive impact of GenAI in boosting firm entry. 

Finally, we also contribute to the booming literature exploring the broad implications of AI on academic research \citep{korinek2023generative, zhu2024divergent}, financial market \citep{cong2025textual, croom2025interactivity, ashraf2025does,xue2025generative}, innovation \citep{wu2025innovation, wang2025artificial}, and international trade \citep{antoniades2025learning}. Our study focuses on the impact of GenAI on entrepreneurship and startup, the engine of economic growth. 

The remainder of this paper proceeds as follows. 
\Cref{sec:background} provides background on ChatGPT and the development of GenAI in China.
\Cref{sec:data} describes the firm registration and patent data, the construction of our fine-grained geo-coded panel as well as the summary statistics.
\Cref{sec:empirical} outlines our empirical strategies, and
\Cref{sec:results} presents the main results and explores mechanisms.
\Cref{sec:test} discusses the placebo and robustness tests.
\Cref{sec:conclusion} concludes.

\section{Release of ChatGPT and the Rise of AI Activity in China}
\label{sec:background}

The public release of ChatGPT by OpenAI in November 2022 represented a global turning point in the diffusion of GenAI. Unlike earlier waves of automation confined to coding or data analytics, ChatGPT demonstrated that large language models (LLMs) could perform a broad spectrum of cognitive and creative tasks—including text generation, document drafting, coding assistance, marketing content, and basic customer interaction—at negligible marginal cost. Its intuitive interface via natural languages and rapid performance improvements signaled a step-change in the accessibility of advanced AI. Within months, ChatGPT surpassed 100 million users worldwide and dominated global media attention, generating an unprecedented awareness shock that transformed how entrepreneurs and small firms perceived the cost and feasibility of launching new ventures. A growing cohort of “AI-native” start-ups built and run by "tiny teams"---often just a handful of engineers using tools like ChatGPT—are already scaling to millions of users and substantial revenue with far less labor and capital than previous generations, illustrating how GenAI can radically change the cost and organizational structure of building new firms.\footnote{https://www.nytimes.com/2025/02/20/technology/ai-silicon-valley-start-ups.html}

The diffusion effects were particularly visible in China, major technology firms rapidly accelerated domestic LLM development.\footnote{Although ChatGPT itself was officially inaccessible in mainland China, the technological and informational spillovers were immediate, disseminated through media coverage and professional networks. Within weeks of its release, Chinese technology firms, developers, and professional users accessed the model through API integrations, enterprise accounts registered abroad, and proxy connections. Moreover, domestic platforms rapidly introduced close substitutes such as Baidu’s ERNIE Bot and Alibaba’s Tongyi Qianwen (Qwen), which replicated many core functionalities of ChatGPT. As a result, both the awareness and practical use of GenAI tools spread widely across China’s entrepreneurial ecosystem.} Between early 2023 and mid-2024, Baidu launched ERNIE Bot\footnote{https://ernie.baidu.com}, Alibaba introduced Tongyi Qianwen (Qwen)\footnote{https://qwen.ai/home}, iFlytek released SparkDesk\footnote{https://xinghuo.xfyun.cn}, Tencent rolled out Hunyuan\footnote{https://hunyuan.tencent.com}, and ZhiPu AI debuted ChatGLM\footnote{https://chatglm.cn}, as well as DeepSeek\footnote{https://www.deepseek.com/en} that has grabbed global headlines. Most of these systems are released as an open-source or open-weight models. This wave of indigenous innovation coincided with a surge of public interest, experimentation, and investment in GenAI applications for marketing, e-commerce, education, and professional services. For self-employed individuals and small entrepreneurs, these tools could dramatically reduce start-up costs by substituting for specialized labor (e.g., designers, coders, or translators) and by enabling professional-grade outputs without prior technical expertise. Collectively, the ChatGPT shock thus acted as a global productivity catalyst—lowering entry barriers and stimulating new firm creation across sectors and regions.

\section{Data and Variables}
\label{sec:data}

We use the administrative firm registration data to measure new firm entry and patent data to proxy AI human capital, and we construct fine-grained geo-coded grids based on firm and patent addresses. This section describes our data sources, sample construction, and main variables along with their corresponding summary statistics. Variable definitions and additional details including their construction are provided in \Cref{sec:appendix-a}.

\subsection{Data Source and Sample Construction}
\label{subsec:data}

\paragraph{Firm Registration Data.}
We obtain firm registration data from the State Administration for Industry and Commerce (SAIC) of the People's Republic of China. The dataset provides detailed firm-level information, including the company name, registered address, contact information, legal representative, industry classification, firm description, registered capital, registration date, operating status, as well as information on executives and both firm and individual shareholders.\footnote{The SAIC dataset has been recently used by \cite{shi2020tiered,allen2024centralization,allen2025ownership,Cai_Gu_Zhao_Zhu_2025}.}

Our sample covers all firms registered between January 2021 and December 2024. We exclude individual industrial and commercial households and remove duplicate entries. The resulting dataset contains 12,820,211 newly established firms. 
Following China's Company Law (2014), we classify a firm as ``small'' if its registered capital is below RMB~1~million and ``large'' otherwise. This yields 5,536,247 small firms and 6,562,515 large firms in our sample.\footnote{Among the 12,820,211 newly registered firms, 721,449 (5.6\%) have missing information on registered capital. As a result, the total number of firms does not equal the sum of small and large firms.} Before ChatGPT released in November 2022, a total of 5.95 million new firms were established, including 2.01 million  (33.8\%) small firms and 3.62 million  (60.8\%) large firms.
In the post-ChatGPT period, new firm registrations rose to 6.87 million, driven by a sharp increase in small-firm entries to 3.53 million (51.4\%), while large-firm entries declined to 2.94 million (42.8\%).

\paragraph{Patent Data and AI Classification.}

To measure AI-related innovation in the pre-ChatGPT period and to proxy AI human capital, we obtain patent data from the China National Intellectual Property Administration (CNIPA) and focus on invention patents filed between 2010 and 2019. Following the classification approach of \citet{fang2025ai}, we apply the AIPatentSBerta model, a transformer-based algorithm pretrained on the full corpus of U.S. patent documents and fine-tuned on labeled Chinese AI and non-AI patents, to identify AI-related patents. This procedure yields 340,771 AI-related invention patents over the period, which we use to construct measures of local AI innovation intensity and AI-specific human capital.


\paragraph{Spatial Grids and Panel Construction.}

To spatially index both patents and firm registrations, we construct a fine spatial grid covering mainland China using the H3 geospatial indexing system, a hierarchical framework that tessellates the Earth into hexagonal grid cells with uniform area and adjacency properties.\footnote{For details, see \href{see details}{https://h3geo.org/}. We use resolution 7, which corresponds to hexagons with a radius of approximately 2 km and an area of roughly 5 km$^2$, comparable to a 2$\times$2 km square.} China spans approximately 18°N–54°N latitude and 73°E–135°E longitude; applying the H3 resolution-7 grid over this range yields 166{,}156 unique hexagonal cells.

Each firm in our registration dataset is geolocated based on its registered firm address, with latitude and longitude coordinates retrieved from Baidu Maps. Similarly, we extract the geographic location of each patent assignee. We then assign both firms and patents to their respective H3 grid cells, indexed by \( g \). For each cell, we compute the total number of AI-related patents filed between 2010 and 2019, denoted by \( \mathit{AIpat}_{g} \), capturing the intensity of AI innovation activity prior to the emergence of GenAI.

Based on the spatial grid and the spatial indexing of firms and patents, we
construct a grid-by-quarter panel spanning 2021Q1 to 2024Q4. This allows us to consistently relate pre-existing AI human capital, as proxied by $\mathit{AIpat}_{g}$, to subsequent patterns of firm formation at a fine spatial resolution.

\subsection{Main Variables}
\label{subsec:main-variables}

Our analysis is conducted at the H3 grid cell \( g \) level. The key variables are: (i) local firm entry, measured by the number of new firm registrations in grid \( g \) and quarter \( t \); (ii) AI-specific human capital, proxied by pre-ChatGPT AI patenting activity in grid \( g \); (iii) entrepreneurial characteristics, including indicators for serial entrepreneurs; and (iv) organizational structure at founding, captured by the number of shareholders and executive team members.

\paragraph{New Firm Creation.} 
Our main dependent variable is the number of new firm registrations in each grid and quarter. For each grid–quarter, we calculate both the total number of newly registered firms and subcategories by firm size.


\paragraph{Pre-ChatGPT Grid-Level AI Human Capital.}
Our identification strategy hinges on the fact that human capital plays a central role in the diffusion and adoption of AI technologies \citep{babina2023firm, gofman2024artificial, andreadis2025local,zhu2024divergent}. To proxy pre-ChatGPT grid level AI-relevant human capital, we use the number of AI-related invention patents filed between 2010 and 2019 in H3 grid cell \( g \), denoted by \( \mathit{AIpat}_{g} \), based on the classification in \Cref{subsec:data}. This measure captures the ex-ante local stock of AI-specific technological capability and serves as our main measure of AI human capital. Our hypothesis is that areas with higher pre-existing AI expertise are better positioned to absorb and deploy GenAI, with entrepreneurs and workers in these regions more likely to recognize, experiment with, and incorporate GenAI into new firm formation.

For our baseline regression analysis, we construct a binary exposure indicator \( \mathit{HighAI}_{g} \), equal to 1 if $\mathit{AIpat}_{g}>0$, i.e., grid \( g \) has at least one AI-related patent in 2010–2019, and 0 otherwise. Under this definition, 6.1 percent of grids (10{,}183 out of 166{,}156) are classified as high AI exposure. We assess robustness to alternative exposure definitions in \Cref{subs:robust_checks}.

\paragraph{Serial Entrepreneur.}

To examine heterogeneity in founder experience, we construct a measure of serial entrepreneurship as a proxy for local entrepreneurial capital. For each newly registered firm, we trace the legal representative across historical records to determine whether the same individual had founded any other firm in the preceding three years. A match indicates prior entrepreneurial experience, classifying the individual as a \emph{serial entrepreneur}.

We then aggregate this information to the grid–quarter level and define \textit{Pct Serial Entrep} as the share of new firm founders in a given cell and quarter who are serial entrepreneurs. To investigate differential patterns by firm size, we separately compute this share for small and large firms, denoted as \textit{Pct Serial Entrep Small} and \textit{Pct Serial Entrep Large}, respectively. These measures allow us to examine whether the rise in firm formation associated with GenAI is driven by novice founders or those with prior experience navigating the startup process.

\paragraph{Organizational Structure at Entry.}
To characterize the structure of firms at entry, we extract detailed information from registration records on founding teams. Specifically, for each newly registered firm, we record: (i) the number of shareholders, (ii) the share of individual (i.e., natural-person) shareholders, and (iii) the number of listed executive members. These metrics respectively capture organizational complexity, ownership composition, and managerial team size.

We aggregate these variables to the grid–quarter level by taking simple averages across all newly registered firms in a given cell and quarter. The resulting panel allows us to analyze how firm-level organizational structure varies across grid and time. In extended analyses, we use these measures to examine whether and how GenAI changes how firms are structured at entry—potentially enabling leaner startups.

\subsection{Descriptive Statistics}

\begin{figure}[htbp]
    \centering
    \caption{Quarterly Trends in Average New Firm Formation by Grid}
    
    \caption*{This figure shows the quarterly trend in new firm formation. The x-axis indicates the quarter, with 2022Q4 marking the release of ChatGPT. The blue solid line represents the average number of new small businesses across all grids in each quarter, while the red dashed line represents the corresponding average number of new large businesses. 
    }

        \centering
        \includegraphics[width=.8\textwidth]{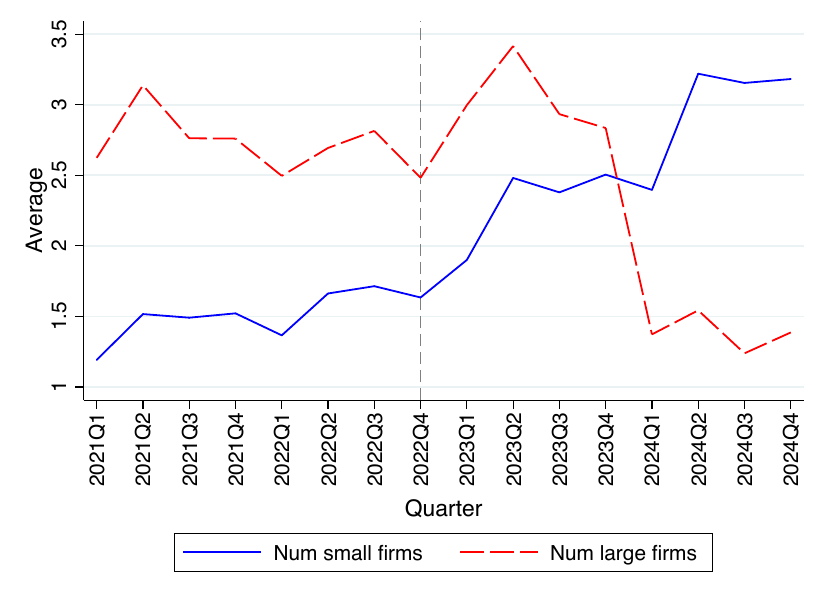}



    \label{fig:trend_firm_entry}
\end{figure}

Table~\ref{tab:summary_statistics} reports summary statistics for the main variables used in our analysis. The sample covers 16~quarters from 2021Q1 to 2024Q4 and includes all H3 grid–quarters with at least one firm registration in our sample period, resulting in a balanced panel of approximately 2{,}658{,}496 observations. On average, each grid hosts around 4.8 new firms per quarter, with substantial spatial heterogeneity. Roughly 46\% of the new entrants are small firms with registered capital below RMB~1~million, while large firms account for the remaining 54\%. The average grid records 2 new small firms and 2 new large firms per quarter (mean value 2.1 and 2.5 respectively), indicating that entry is not dominated by micro‐enterprises. Figure~\ref{fig:trend_firm_entry} further reports the quarterly trends of average new firm formation per grid. The figure shows that prior to 2022Q4, the average number of new small firms per grid is stable at around 1.4–1.7 per quarter, while large firms average about 2.6–3.0. After 2022Q4, small firm entry increases to roughly 2.5–3.2 per quarter, whereas large firm entry declines to about 1.3–1.5. This pattern indicates a structural shift in the composition of new firm formation toward smaller entrants.

For variables related to entrepreneur characteristics, 
we have to restrict attention to grid–quarter observations with at least one new firm.
Consequently, the usable sample size declines by around 60.1\% to 1.04 million. Summary statistics indicate that the prevalence of serial entrepreneurs is modest: the mean value of \textit{Pct Serial Entrep} is 26.7\%, suggesting that just over one in four new founders have previous firm-creation experience. For small firms, \textit{Pct Serial Entrep Small} averages slightly lower at 16.3\%, consistent with a greater presence of first-time entrepreneurs in smaller ventures.

The average number of shareholders per firm is 1.5 (1.4 among small firms), indicating that start-ups are typically founded by a very limited number of equity holders. Correspondingly, individual shareholders account for 93.2\% of the total on average, and an even higher share of 96.1\% among small firms, implying that most start-ups are closely held by a few private owners. The mean size of the executive team is about two people, with limited variation across firm size categories, suggesting that founding teams tend to remain relatively small in both small and larger start-ups.



\begin{figure}[htbp]
    \centering
    \caption{Histogram Distribution of AI Patent}

    \caption*{This figure shows the distribution of the number of AI patents across grids that have at least one AI patent.}

    \begin{subfigure}[t]{0.8\textwidth}
        \centering
        \includegraphics[width=\textwidth]{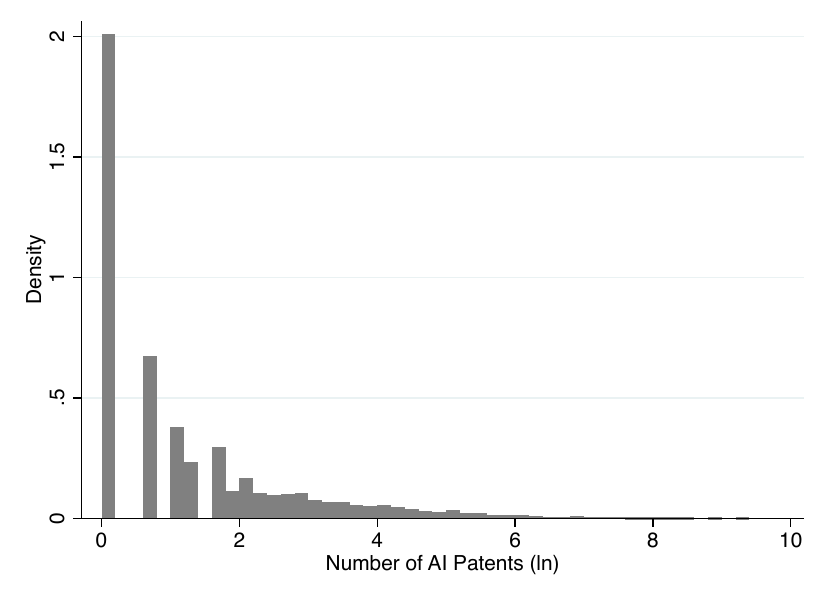}
    \end{subfigure}

    \label{fig:aipatent_hist}
\end{figure}

Turning to the key explanatory variable, AI exposure exhibits pronounced spatial variation and concentration. Figure~\ref{fig:aipatent_hist} shows that the distribution of \textit{AIpat\_g} is highly right-skewed, with a mean of 2.1 and a median of 0. Most grids host only a handful of AI patents, while a small subset exhibits much higher patent intensity, leading to a long right tail and indicating strong spatial concentration of AI innovation. Under our baseline definition, approximately 6\% of grids are classified as high-AI regions (\textit{HighAI\_g}=1). These high-AI grids are primarily located in major metropolitan areas such as Beijing, Shanghai, Shenzhen, and Hangzhou.

Overall, the summary statistics confirm that the sample captures substantial variation in firm entry, entrepreneurial composition, and local AI exposure across grid and time. The data also show that regions with stronger pre-existing AI activity tend to be more urban and economically developed, a pattern we explicitly control for in the regression analysis.

\begin{center}
[Insert \Cref{tab:summary_statistics} Here]

[Insert \Cref{fig:trend_firm_entry} Here]

[Insert \Cref{fig:aipatent_hist} Here]
\end{center}

\paragraph{Geographic Concentration and Heterogeneity of AI Innovation Activity.}

We document the spatial distribution of pre-ChatGPT AI patenting. Across grids, we show both regional variation and diversification, as well as pronounced within-city heterogeneity in AI patents. These patterns provide the foundation for our identification strategy, in which local AI patenting serves as a proxy for AI-specific human capital.

\begin{figure}[htbp]
    \centering
    \caption{Geographical Distribution of AI Patents and Inventors}
    \label{fig:overall_geographical_distribution}
        \centering
        \includegraphics[width=\textwidth]{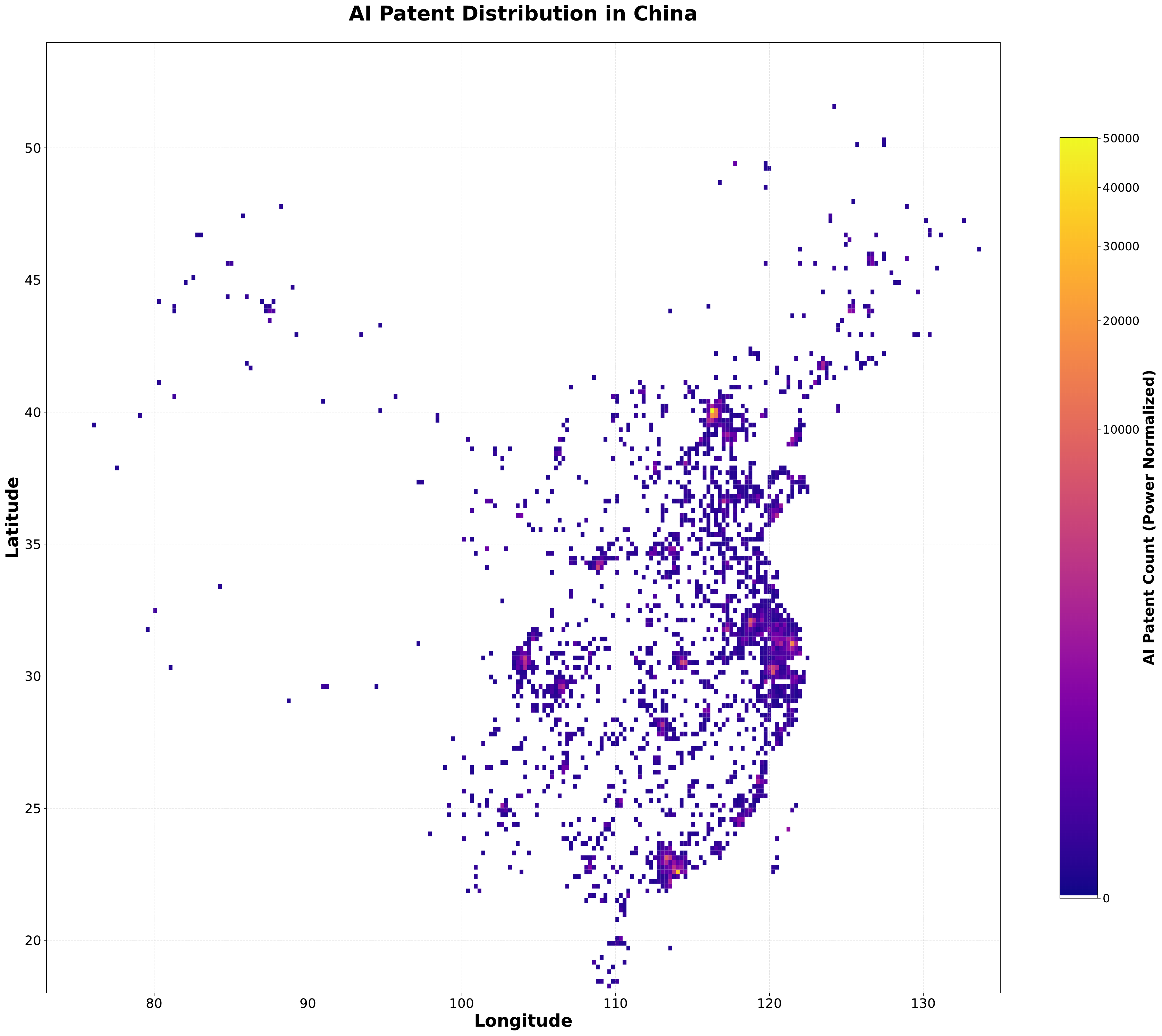}

    \label{fig:aipatent_dist}
\end{figure}

Figure~\ref{fig:overall_geographical_distribution} visualizes the spatial distribution of AI-related patents and inventors across China at the H3 grid-cell level. Each point represents a 2km-resolution hexagonal cell, with intensity power-normalized to reflect the local concentration of AI activity. As expected, both AI patents and inventors are concentrated in major innovation hubs such as Beijing, Shanghai, Shenzhen, Hangzhou, and Guangzhou. We also find the following features.

First, although AI innovation is disproportionately concentrated in China's southeastern corridor, it is not narrowly confined to top-tier cities. We observe meaningful \textit{regional diversification}: inland growth poles and homes to dense cluster of high-education institutions such as Chengdu, Xi’an, Wuhan, and Hefei, as well as many second- and third-tier cities, exhibit substantial AI activity. Overall, the spatial spread of AI patents and inventors aligns closely with the economically active zone \textit{east of the Hu Line} (also known as the Heihe–Tengchong Line), a well-known geographic demarcation separating China's densely populated and industrialized eastern region from its sparsely populated west.\footnote{The Hu Line traces a line from Heihe in the northeast to Tengchong in the southwest. It roughly divides China into a highly urbanized, economically developed eastern region (comprising only around 40\% of land area but home to around 94\% of the population) and a less-developed western hinterland.}

Second, and more critically for our identification strategy, there exists substantial \textit{within-city heterogeneity} in AI exposure. Even within the same metropolitan area, certain H3 grids exhibit intense innovation activity, while nearby grids remain largely inactive. 
To illustrate this variation, we zoom in on three representative innovation hubs—Beijing, Shanghai, and Shenzhen—and plot the spatial distribution of AI patents at the grid level. 
As shown in Figure \ref{fig:beijing-shanghai-shenzhen}, even within these leading technology centers, some neighborhoods exhibit dense clusters of AI patenting prior to the GenAI era, while adjacent grids display little to no AI activity. 
This fine-grained variation provides the foundation for our empirical strategy: it allows  us to flexibly absorb unobserved city-level shocks while retaining variation in AI activity, and it serves as a valid source of cross-grid variation in AI human capital.
By comparing grids that share similar institutional, regulatory, and macroeconomic environments but differ sharply in pre-existing AI intensity, we can identify the local entrepreneurial response to the diffusion of GenAI.

\begin{center}
[Insert \Cref{fig:overall_geographical_distribution} Here]
\end{center}

\section{Empirical Strategy}
\label{sec:empirical}

\subsection{Baseline Specification}\label{sec:baseline}
As detailed in \Cref{subsec:data}, we construct a quarterly panel at the grid-cell level, measuring the number of newly registered firms in each hexagonal H3 cell. To estimate the effect of GenAI on firm formation, we exploit cross-sectional variation in pre-2020 exposure to AI-specific human capital, as captured by the intensity of AI patents filed by firms or inventors within each grid. Grids with high concentrations of AI patents serve as treated units, while those with low or no prior exposure form the control group.  

The November 2022 release of ChatGPT provides a sharp and plausibly exogenous technological shock that dramatically increased public awareness and experimentation with GenAI tools. Our empirical strategy compares changes in firm formation between high- and low-AI (exposure) grids before and after this event, relying on fine-grained \emph{within-city} variation to isolate local treatment effects. The baseline difference-in-differences specification is as follows:
\begin{equation}
\label{eq:did}
Y_{gt}
=
\beta\big(\mathrm{Post}_{t} \times \mathrm{HighAI}_{g}\big)
+\mu_{g\times q(t)}
+\lambda_{c(g)\times t}
+\varepsilon_{gt},
\end{equation}
where \(Y_{gt}\) denotes the number of newly registered firms in grid \(g\) at time \(t\); \( \mathrm{HighAI}_g \) equals one for grids with high pre-2020 AI patenting intensity; and \( \mathrm{Post}_t \) equals one for quarters following ChatGPT’s release (2022Q4 onward).
The function $q(t)$ maps time $t$ into its calendar quarter, and $c(g)$ denotes the city containing grid $g$.
The coefficient of the interaction term \( \beta \) measures how much more firm creation changes in AI-intensive grids relative to less exposed ones within the same city after ChatGPT's release.

The specification includes both \emph{grid-by-calendar-quarter} fixed effects (\(\mu_{g\times q(t)}\)) and \emph{city-by-quarter} fixed effects (\(\lambda_{c(g)\times t}\)). This high-dimensional fixed-effect structure absorbs (i) persistent grid-specific seasonality and any time-invariant unobservables within each grid, and (ii) any time-varying city-level shocks, such as local policy initiatives, changes in the business environment, industrial restructuring, or macroeconomic fluctuations. Standard errors are clustered at the city level. Identification thus arises exclusively from \emph{within-city, within-quarter} differences across neighboring grids that differ in pre-existing AI exposure.

This spatially granular design offers several identification advantages. By comparing grids within a common institutional and economic context, we eliminate bias from unobserved regional heterogeneity and policy differences across cities. The grid-level fixed effects control for micro-spatial factors such as persistent differences in local infrastructure or land use, while the city-by-quarter fixed effects purge all contemporaneous shocks at the city level. The global and unanticipated nature of the ChatGPT release strengthens the quasi-experimental setting, providing a sharp and plausibly exogenous timing for treatment. 

To evaluate remaining identification concerns, we perform event-study tests for parallel pre-trends in \Cref{subsec:event_study} and a battery of robustness and placebo tests in \Cref{sec:test} including placebo regressions using non-AI or residualized patent measures, and random reassignment of treatment labels, all of which consistently yield null or attenuated effects. These exercises confirm that the estimated coefficients capture the causal impact of GenAI through localized AI-specific human capital, rather than reflecting general entrepreneurial pattern or spurious spatial correlations.

\subsection{Heterogeneous Effects by Firm Size}

As we argue, GenAI, as a general-purpose technology, redefines the economics of entrepreneurship. 
It automates language-based information-processing (reasoning, synthesis, coding, and content generation), alongside creative, managerial, and operational tasks, thereby lowering the barriers to capabilities that were once capital-intensive or concentrated among specialized labor and well-resourced organizations.
In practice, GenAI allows a very small founding team, or even a single entrepreneur, to handle functions that previously required multiple specialists in product development, marketing, operations, and customer support.
This transformation shifts the production frontier of what lean ventures can achieve and challenges traditional assumptions about firm size, scale, and the division of labor.
A growing wave of \emph{GenAI-native} firms—such as \textit{Midjourney}, \textit{Cursor}, and \textit{Perplexity}—illustrates this paradigm shift. These startups achieve rapid growth with minimal staff and modest capital by embedding generative models directly into design, engineering, and marketing workflows. 
In this sense, GenAI operates as a “digital co-founder,” lowering the minimum viable scale of new ventures. 

Building on this intuition, we hypothesize that GenAI disproportionately lowers entry barriers for small and resource-constrained firms relative to larger incumbents. The mechanism rests on differences in fixed costs and internal capabilities at the early stages of firm formation. Traditional firm creation requires access to specialized expertise, such as in software development, marketing, legal, and operational management, creating a fixed cost structure that weighs most heavily on startups. GenAI relaxes these constraints by automating or augmenting mange of these tasks, allowing small teams or individual entrepreneurs to substitute AI capabilities for costly human capital.

This mechanism is consistent with broader theoretical insights from the entrepreneurship literature. Specifically, in standard models of occupational choice under frictions (e.g., \citealp{evans1989estimated,kerr2015financing}), reductions in entry costs expand the extensive margin of entrepreneurship by allowing financially or skill-constrained agents to overcome fixed investment thresholds. If GenAI compresses these thresholds, we should observe  greater firm entry in regions with higher effective access to GenAI, particularly among smaller firms and first-time entrepreneurs. 

We test this prediction empirically by estimating heterogeneous treatment effects by firm size.
Specifically, we extend our baseline difference-in-differences specification \ref{eq:did} to separate small and large entrants:
\begin{equation}
\label{eq:did-size}
Y_{gt}^{\text{size}} = \beta^{\text{size}} \left( \mathrm{Post}_t \times \mathrm{HighAI}_g \right)
+ \mu_{g\times q(t)} + \lambda_{c(g)\times t} + \varepsilon_{gt}^{\text{size}},
\end{equation}
where $Y_{gt}^{\text{size}}$ the number of new firms of a given size category (small or large) in grid $g$ and time $t$. 
We classify firms as small or large based on the RMB 1 million registered-capital cutoff described in \Cref{subsec:data}, and we report robustness checks using alternative capital thresholds in \Cref{subsec:threshold}.
If GenAI primarily reduces fixed costs or substitutes for managerial labor, we would expect \( \beta^{\text{small}} > 0 \) and \( \beta^{\text{large}} \leq 0 \).

\subsection{Dynamic Model}
\label{subsec:event_study}

To examine the temporal evolution of treatment effects and validate the parallel trends assumption, we estimate an event-study specification of the following form:

\begin{equation}
\label{eq:event}
Y_{gt}
=
\sum_{k \neq -1} \beta_k \cdot {1}\{t - t_0 = k\} \times \mathrm{HighAI}_g
+ \gamma_g
+ \lambda_{c(g)\times t}
+ \varepsilon_{gt},
\end{equation}
where \( t_0 \) denotes the reference quarter (2022Q4), and each coefficient \( \beta_k \) captures the difference-in-differences-style effect: the difference in outcomes between high- and low-AI exposure grids \( k \) quarters away from the baseline, relative to the reference period \( k = -1 \) (the quarter immediately preceding ChatGPT's release). Grid fixed effects \( \gamma_g \) control for time-invariant heterogeneity across locations, while city-by-quarter fixed effects \( \lambda_{c(g)\times t} \) flexibly absorb common time-varying shocks at the city level.

\begin{figure}[htbp]
    \centering
    \caption{Dynamic Effect of the Release of ChatGPT on Firm Creation}
    
    \caption*{This figure illustrates the dynamic evolution of new firm formation in the quarters surrounding the release of ChatGPT. The X-axis represents the quarter relative to the release of ChatGPT, where 0 corresponds to 2022Q4. The empirical specification is based on regression Equation~\eqref{eq:event}. Panels A–C display results for the following dependent variables: (A) the number of new firms, (B) the number of small firms, and (C) the number of large firms. The Y-axis plots the estimated coefficients, $\beta_k$.}

    \begin{subfigure}[t]{0.8\textwidth}
        \centering
        \includegraphics[width=\textwidth]{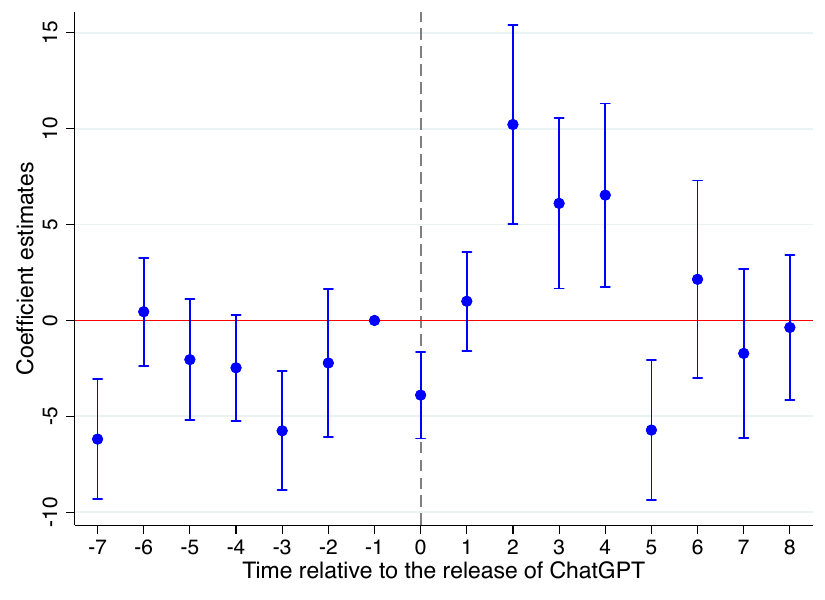}
        \caption{Panel A: Entry of New Firms}
        \label{fig:dynamic-panelA}
    \end{subfigure}

    \vspace{1em} 

    \begin{subfigure}[t]{0.45\textwidth}
        \centering
        \includegraphics[width=\textwidth]{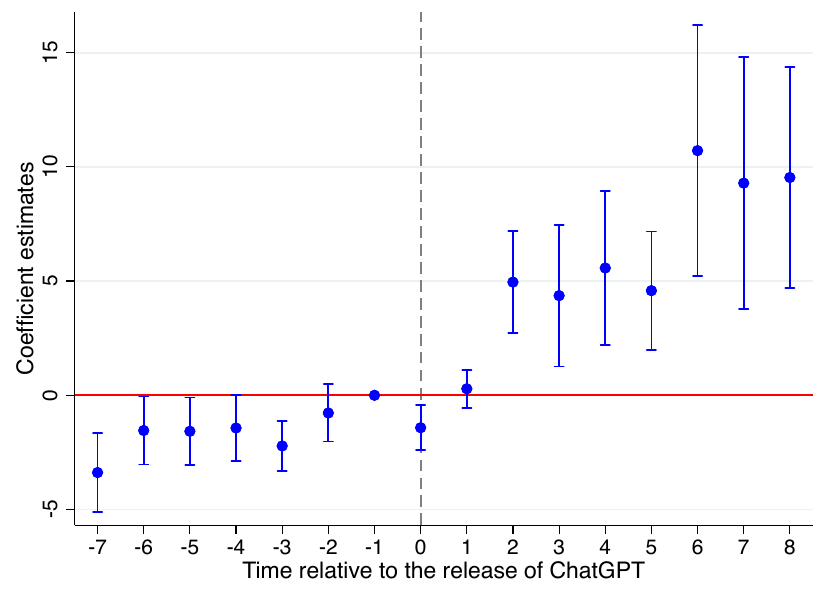}
        \caption{Panel B: Entry of Small Firms}
        \label{fig:dynamic-panelB}
    \end{subfigure}
    \hfill
    \begin{subfigure}[t]{0.45\textwidth}
        \centering
        \includegraphics[width=\textwidth]{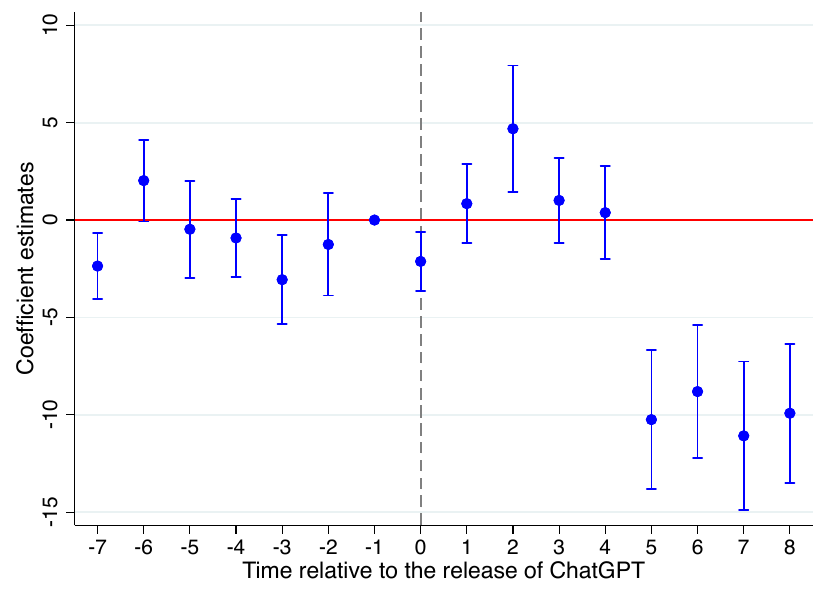}
        \caption{Panel C: Entry of Large Firms}
        \label{fig:dynamic-panelC}
    \end{subfigure}

    \label{fig:dynamic_gpt_firm_entry}
\end{figure}

We plot the estimated event-time coefficients \( \widehat{\beta}_k \) with 99\% confidence intervals in \Cref{fig:dynamic_gpt_firm_entry}. Under the parallel trends assumption, pre-treatment estimates (\( k < 0 \)) should be statistically indistinguishable from zero, while significant post-treatment effects (\( k \geq 0 \)) would indicate a divergence in firm formation between high- and low-exposure grids following the release of GenAI technologies.

\begin{center}
[Insert \Cref{fig:dynamic_gpt_firm_entry} Here]
\end{center}

\section{Empirical Results and Mechanisms}
\label{sec:results}

\subsection{Baseline results}
 
Table~\ref{tab:regression_results} presents the baseline difference-in-differences estimates of the impact of genAI on new firm creation based on \Cref{eq:did,eq:did-size}. Across all specifications, the coefficient on the interaction term \textit{Post} $\times$ \textit{HighAI} is positive and statistically significant, indicating that firm entry increased more strongly in grids with high pre-existing AI activity following the release of ChatGPT in November 2022. The magnitude of the effect is economically meaningful.
Column 1 shows the results based on \Cref{eq:did} (controlling for
grid-by-calendar-quarter and city-by-quarter fixed effects), and the estimated coefficient on \textit{Post} $\times$ \textit{HighAI} is 5.038 (s.e.=1.395), significant at the 1\% level. This suggests that, all else equal, high-AI grids experienced an average increase of about five additional new firms per grid-quarter following the introduction of ChatGPT. Aggregating across all 10,183 high-AI grids and 8 quarters after the release of ChatGPT, this effect implies approximately \(5.038 \times 10{,}183 \times 8 \approx 0.41 \) million additional firm entries post ChatGPT, accounting for around 6.0 percent of the total 6.87 million new firms established nationwide during the same period.



Columns~2-3 separate the estimates by firm size based on \Cref{eq:did-size}. Consistent with the hypothesis that GenAI primarily reduces fixed costs and substitutes for scarce managerial labor, the positive effect is entirely driven by small firms. The estimated coefficient for small-firm entry is positive and significant, whereas the corresponding coefficient for large firms is negative and significant. Correspondingly, the coefficient on the interaction term for small firms is 7.704, suggesting that all else equal, high-AI grids experienced an average increase of approximately eight additional firms per quarter. This divergence suggests that GenAI tools, by automating core tasks like marketing, customer interaction, and software development, compress the scale required to operate a business. Entrepreneurs who might previously have needed larger teams and capital to launch may now find it feasible to enter as smaller, more agile firms. The decline in large firm entry in high-AI grids may reflect this substitution toward leaner organizational forms, rather than a decline in overall entrepreneurial activity. Since all regressions exploit within-city variation and absorb city-level shocks, the results are not confounded by local policy interventions or macroeconomic trends. Instead, they capture how heterogeneous AI exposure shapes local responses to a frontier general-purpose technology. Overall, these results suggest that the diffusion of GenAI disproportionately stimulated entrepreneurship among small and resource-constrained firms, supporting the view that GenAI functions as a ``digital cofounder'' that lowers entry barriers.

\begin{center}
[Insert \Cref{tab:regression_results} Here]
\end{center}

Table~\ref{tab:regression_results_winsor} reports robustness checks that winsorize the dependent variable to mitigate the influence of extreme values. Columns~1–3 winsorize within quarter, while Columns 4-6 winsorize across the full sample. Across all specifications, the interaction term \( \textit{Post}_t \times \textit{HighAI}_g \) remains statistically significant and economically sizable. For total new firm creation, the coefficient is 3.692 (s.e.\;=\;0.826) in Column~1 and 2.518 (s.e.\;=\;0.667) in Column 4, suggesting increase of approximately three to four additional firms (52.2 to 76.5 percent relative to the sample mean of new firms. 

Columns~2–3 and 5–6 report results separately by firm size. Consistent with the mechanism highlighted in Table~\ref{tab:regression_results}, the post-ChatGPT rise in firm entry is concentrated among small firms: the coefficients are 6.170 (s.e.\;=\;0.800) with within-quarter winsorization and 4.911 (s.e.\;=\;0.528) with full-sample winsorization. By contrast, large-firm entry declines: the coefficients are \(-2.928\) (s.e.\;=\;0.332) and \(-2.922\) (s.e.\;=\;0.336). Taken together, these robustness checks confirm that the baseline effects are not artifacts of outliers or specific fixed-effects choices: GenAI is associated with a sizeable increase in small-firm formation and a corresponding decline in large-firm entry within the same city and quarter.


\begin{center}
[Insert \Cref{tab:regression_results_winsor} Here]
\end{center}

\subsection{Industry Heterogeneity}

Table~\ref{tab:industry-heterogeneity} reports industry-specific difference-in-differences estimates of the interaction term \textit{Post} $\times$ \textit{HighAI}, examining how the impact of GenAI on firm entry varies across industries. The specification is identical to our baseline model \eqref{eq:did} but estimated separately by industry.
Panel A reports the top fifteen industries ranked by the estimated coefficients. The results reveal substantial heterogeneity in the responsiveness of firm entry to GenAI. The largest effects are observed in the \textit{Retail}, \textit{Business Services}, and \textit{Technology Promotion and Application Services} sectors, with coefficients of 1.625 (s.e.\;=\;0.545), 0.977 (s.e.\;=\;0.292), and 0.871 (s.e.\;=\;0.231), respectively, all statistically significant at the 1\% level. These magnitudes indicate that grids with higher pre-existing AI activity experience stronger post-ChatGPT increases in firm formation within commercially oriented and service-intensive industries.

Several additional sectors also exhibit positive and significant coefficients, including \textit{Wholesale}, \textit{Entertainment}, \textit{Catering}, and \textit{Culture and Arts}, suggesting that GenAI facilitated new firm creation in customer-facing and creative domains where automation and content generation tools directly augment production. 
By contrast, Panel B reports the bottom fifteen industries ranked by the estimated coefficients where traditional and capital-intensive industries including construction and manufacturing exhibit small or even negative effects.
These patterns reinforce the interpretation that the entrepreneurial impact of GenAI operates primarily through demand-side adoption and creative applications. Overall, \Cref{tab:industry-heterogeneity} underscores that the sectors whose core of business activities rely on knowledge work, creativity, marketing, and digital service are most responsive to GenAI.

\begin{center}
[Insert \Cref{tab:industry-heterogeneity} Here]
\end{center}

\paragraph{Sectoral heterogeneity in AI-Relevance Scores.} 
We further exploit the rich information in the firm registration data to study industry heterogeneous effects. Since GenAI affects industries differently, we first use the official 96 industry labels to classify firms by sector. To obtain a more AI-specific measure of exposure at sectoral level, we then analyze firms’ business descriptions: we extract keywords using topic methods and adopt a state-of-the-art language model to assign three continuous scores to each firm: an \emph{AI-upstream} score (the extent to which the firm’s activities relate to upstream AI, such as foundation model development, data infrastructure, or computing resources), an \emph{AI-downstream} score (the extent to which the firm operates in downstream AI uses, such as application development, product integration, or end-user services) and an \emph{AI entrepreneurship helpfulness} score, which captures the degree to which GenAI can assist and support a firm's core business activities or lower key entrepreneurial frictions, for example by speeding up product prototyping and iteration, helping inexperienced founders perform specialized tasks (e.g., basic coding, legal/financial drafting, market research), and facilitating content creation, personal branding, low-cost experimentation, or digital marketing.
All three scores range from $-100$ to $100$, with higher values indicating stronger relevance or complementarity with GenAI.

The construction proceeds in two stages. First, we represent each firm’s business scope as a weighted combination of latent \emph{textual factors} derived from its official business description following the approach by \cite{cong2025textual}. These topics summarize the underlying business activities and 
provide a less noisy representation of its underlying business than the raw text. Second, we employ the state-of-the-art large language model GPT-4o \citep{hurst2024gpt} to evaluate each topic’s relevance to AI across multiple dimensions. Each firm’s AI relevance score is then computed as a weighted average of these topic-level evaluations, where the weights correspond to the firm’s topic loadings.
We validate the measure by manually reviewing a random subset of firms, i.e., examining their business websites, product descriptions, and public filings, and find a close correspondence between these qualitative assessments and our model-based scores. Full construction details and validation procedures are provided in Appendix~\ref{sec:AI-relevance-score}.

As Table \ref{tab:summary_statistics} shows, around 20\% of the new entrants have high (compared to low) AI upstream scores; 52\% of the new entrants have high (compared to low) AI downstream scores; 75\% of the new entrants have high (compared to low) AI entrepreneurship scores. By examining how the effects vary across sectors along these three AI-related dimensions, we characterize industry-level heterogeneity in exposure to GenAI.

Table~\ref{tab:industry} presents the results. Columns~1–4 use alternative dependent variables: the number of high-upstream, low-upstream, high-downstream, and low-downstream new firms in each grid–quarter, respectively, where “high” denotes a score $>0$ and “low” a score 
$\leq 0$. Columns 5–6 similarly use as dependent variables the number of new firms with high versus low AI-entrepreneurship helpfulness (score $>0$ vs. $\leq 0$).
Across specifications, the interaction term \textit{Post} $\times$ \textit{HighAI} is positive and statistically significant in AI-relevant firms. The coefficients are much larger for low-upstream (4.199 with s.e.\;=\;1.225) and high-downstream (3.323 with s.e.\;=\;0.789) firms than for high-upstream (0.846 with s.e.\;=\;0.388) and low-downstream (1.722 with s.e.\;=\;0.753) firms, suggesting that the diffusion of GenAI has primarily stimulated entry in industries that adopt or apply AI rather than those that develop it. Upstream industries (such as semiconductors, cloud infrastructures, and data centers) remain capital- and expertise-intensive, limiting new entry even after the release of ChatGPT. By contrast, downstream industries—such as digital marketing, content generation, and business software integration—can more readily embed off-the-shelf generative models into their products and workflows. 

Columns~5–6 examine heterogeneity by AI entrepreneurship helpfulness. The coefficient for high-entrepreneurship sectors is 5.892 (s.e.\;=\;1.302), significant at the 1\% level, while that for low-entrepreneurship industries is negative (\(-0.848\), s.e.\;=\;0.300).
Recall that the entrepreneurship helpfulness score captures the extent to which GenAI can support a firm’s core activities or relax key entrepreneurial frictions.
This contrast between high- and low-entrepreneurship sectors suggests that GenAI most strongly stimulates entrepreneurial activity in sectors that are both technologically close to AI and naturally suited to individual or small-team creative work, 
such as digital marketing, software development, and data analytics/
By contrast, sectors with weaker entrepreneurial or technological complementarities to AI exhibit little or even negative response. Overall, these results reinforce the view that the diffusion of GenAI operates through local technological and human-capital complementarities, amplifying entrepreneurial activity precisely where AI capabilities are most easily integrated into new firm creation.

\begin{center}
[Insert \Cref{tab:industry} Here]
\end{center}

\subsection{Mechanisms}
\label{subsec:mechanisms}

The results above show that the diffusion of GenAI substantially increased new firm formation, particularly among small and resource-constrained entrants. In this subsection, we examine the mechanisms behind this effect. GenAI may stimulate entrepreneurship by relaxing key constraints that traditionally hinder new entry: (i) \textit{experience-related constraints}, as AI tools substitute for the skills, and managerial experience typically accumulated by older or serial founders; (ii) \textit{financial constraints}, by reducing the need to pool capital from multiple shareholders to launch a new firm; and (iii) \textit{labor constraints}, by automating managerial and operational tasks that would otherwise require hiring additional executives. We test these channels using proxies for founders’ prior experience in Table~\ref{tab:serial_entrepreneurship}, 
number of shareholders in Table~\ref{tab:shareholder}, and executive team size in Table~\ref{tab:executive}.

Table~\ref{tab:serial_entrepreneurship} presents evidence on the entrepreneurial experience channel.
The dependent variable is the average share of serial entrepreneurs defined  in \Cref{subsec:main-variables}—legal representatives who had established at least one other firm within the preceding three years—within each grid–quarter. Column~1 reports estimates for the full sample, while Columns~2–3 distinguish small and large firms. Across specifications, the interaction term \textit{Post} $\times$ \textit{HighAI} is negative and statistically significant in the aggregate, indicating that the post-ChatGPT increase in firm entry was primarily driven by first-time entrepreneurs rather than experienced repeat founders. In Column~1, the coefficient of \(-0.405\) (s.e.\;=\;0.201) suggests that the share of serial founders declined by roughly 1.5~percent relative to its pre-ChatGPT mean of 27.5 percent. 

Columns~2–3 further reveal that this decline is concentrated entirely among small firms: the coefficient for small-firm founders is \(-2.456\) (s.e.\;=\;0.266), highly significant, whereas the effect for large firms is positive at 0.648 (s.e.\;=\;0.294). These patterns suggest that GenAI has lowered entry barriers associated with prior entrepreneurial experience, enabling individuals without a founding track record to establish new firms. In high-AI regions, ChatGPT and related tools appear to substitute for experiential knowledge—offering guidance, drafting documents, and automating early-stage tasks that would otherwise require seasoned founders or extensive professional networks. 

\begin{center}
[Insert \Cref{tab:serial_entrepreneurship} Here]
\end{center}

Table~\ref{tab:shareholder} reports estimates for the financing channel, examining whether GenAI reduces the need to pool capital among multiple shareholders at entry. Panel~A focuses on the number of shareholders per new firm. Across all specifications, the coefficient on \textit{Post} $\times$ \textit{HighAI} is negative and statistically significant, indicating that firms founded in high-AI grids after the release of ChatGPT are created with fewer shareholders. In Column~1, the coefficient of \(-0.0211\) (s.e.\;=\;0.00747) implies a decline of roughly 1.4\% relative to the mean number of shareholders (1.51). When splitting by firm size in Columns~2-3, the decline is present for both small and large firms, consistent with the idea that GenAI lowers the initial financing and coordination burden regardless of firm scale. These findings suggest that founders in AI-exposed regions require fewer co-investors or partners to mobilize the minimum viable capital necessary to launch a business.

Panel~B examines whether the composition of shareholders shifts toward individual owners. Across all columns, the estimated effects are small in magnitude and statistically insignificant. The lack of meaningful change in ownership composition indicates that the decline in the number of shareholders is not driven by substitution between individual and corporate shareholders. Instead, the results point to a broader reduction in financial frictions at entry: firms can be launched by fewer equity founders without altering the underlying source of ownership capital. Taken together, Table~\ref{tab:shareholder} supports the mechanism that GenAI relaxes financing constraints by lowering the amount of human and financial capital required to start a firm. This evidence complements the experience channels documented earlier, suggesting that GenAI enables more streamlined and capital-lean firm formation.

\begin{center}
[Insert \Cref{tab:shareholder} Here]
\end{center}

Table~\ref{tab:executive} evaluates the labor substitution channel by examining whether GenAI reduces the number of executive members required to launch a new firm. The dependent variable is the average size of the executive team in each grid–quarter. Across all specifications, the coefficient on \textit{Post} $\times$ \textit{HighAI} is negative and statistically significant, indicating that firms established in high-AI grids after the release of ChatGPT rely on fewer executives at entry. In Column~1, the estimate of \(-0.0161\) (s.e.\;=\;0.00392) represents a decline of around 0.8\% relative to the pre-ChatGPT mean executive team size of 2.032. 

Columns~2-3 reveal that the decline is concentrated among small firms. The coefficient for small-firm executive team size is \(-0.0187\) (s.e.\;=\;0.00401), highly significant, whereas the effect for large firms is small in magnitude and statistically insignificant. These patterns suggest that GenAI substitutes for early-stage managerial labor, enabling new ventures—particularly smaller ones—to operate with leaner founding teams. Along with the declines in serial entrepreneurship (Table~\ref{tab:serial_entrepreneurship}) and the number of shareholders (Table~\ref{tab:shareholder}), the results provide consistent evidence that GenAI relaxes multiple complementary input constraints in firm formation, allowing less-experienced and capital-constrained founders to enter with smaller founding teams and lower organizational complexity.

\begin{center}
[Insert \Cref{tab:executive} Here]
\end{center}

\subsubsection{Serial Entrepreneurs and the Substitution Towards Smaller Firms}

Our baseline analysis shows that the diffusion of GenAI is associated with a significant rise in the entry of small firms alongside a significant decline in the number of newly established large firms. This shift in the composition of new ventures is consistent with a \textit{substitution effect} in entrepreneurial strategy.

One natural interpretation is that GenAI lowers fixed costs and substitutes for creative, information-processing, and managerial labor, enabling entrepreneurs to launch viable businesses with fewer resources as documented in previous section. Even experienced founders—those with the capital and capacity to start larger firms—may instead choose to establish smaller, more agile ventures in the ChatGPT era. By automating key functions such as marketing, customer engagement, and software development, GenAI reduces the minimum efficient scale required to operate a business. In this subsection, we directly test this substitution hypothesis by examining whether serial entrepreneurs downsize the scale of their new ventures following ChatGPT's release, particularly in regions with high pre-2019 AI-patent activity, where local founders possess more AI-relevant human capital and are more likely to effectively leverage GenAI tools.

For each new firm established by a serial entrepreneur, we compare its scale with that of the entrepreneur’s most recent prior venture, using \textit{registered capital}—a standardized and regulatorily disclosed measure of initial firm size—as our proxy for entry scale. We then assess whether the relative firm size of new establishments declines more sharply after ChatGPT’s release in high-AI exposure grids. 
We estimate the same difference-in-differences specification as in Equation~\eqref{eq:did}, but with alternative dependent variables that capture the \emph{relative} scale of new firms founded by serial entrepreneurs in grid 
$g$ and quarter $t$.
Specifically, in \textit{Panel A} of Table \ref{tab:size_shrinkage_for_cumbuersom}, the dependent variable \( Y_{gt} \) denotes the percentage of new firms established by a serial entrepreneur in grid $g$ and quarter $t$ whose registered capital exceeds that of the entrepreneur’s previous firm.
In \textit{Panel B} of Table \ref{tab:size_shrinkage_for_cumbuersom}, \( Y_{gt} \) denotes the  grid-quarter average ratio of registered capital in the new firm established
by a serial entrepreneur in grid $g$ and quarter $t$ to that in the previous firm, trimmed at the top 1\% and bottom 1\% to mitigate outlier influence.  
The interaction term \( \textit{Post}_t \times \textit{HighAI}_g \) captures the differential change in new-firm scale among serial entrepreneurs in AI-intensive grids following ChatGPT’s release.  

Table \ref{tab:size_shrinkage_for_cumbuersom} shows the results. In \textit{Panel A}, the coefficient on \textit{Post} $\times$ \textit{HighAI} is consistently negative and statistically significant, indicating that serial entrepreneurs in high-AI regions tend to reduce firm size after the release of ChatGPT.  The effect is particularly strong among small new firms below the threshold of 1 million RMB: the estimate of $-5.416$ (s.e.\;=\;0.483) in Column~2 implies that the share of new firms whose registered capital exceeds that of the entrepreneur’s previous firm decreases by approximately 5.4 percentage points.
Even for large firms, the coefficient remains negative ($-0.917$, s.e.\;=\;0.241), suggesting a modest but statistically significant tendency toward smaller scale.

\textit{Panel B} corroborates this pattern using the average ratio of new-to-old registered capital as a continuous measure of downsizing.  
The estimated coefficient on the interaction term \textit{Post} $\times$ \textit{HighAI} is $-1.983$ (s.e.\;=\;0.560) for the overall sample, and the magnitude becomes substantially larger for small firms, at $-7.100$ (s.e.\;=\;1.148). 
This result indicates that, following the release of ChatGPT, serial entrepreneurs in high-AI grids started new firms with substantially smaller capital relative to their previous firms, compared with entrepreneurs in low-AI grids. 
On average, new firms are about seven times smaller than the previous ones in high-AI regions.
The negative and statistically significant effect indicates a clear contraction in startup scale among repeat founders during the era of GenAI.  Even among large firms, the estimated coefficient remains negative ($-0.180$, s.e.\;=\;0.023), suggesting that the downscaling tendency extends across the size distribution—albeit more modestly for capital‑intensive ventures.  

\begin{center}
[Insert \Cref{tab:size_shrinkage_for_cumbuersom} Here]
\end{center}

Overall, these findings provide direct evidence of a substitution mechanism: experienced entrepreneurs are recalibrating the optimal scale of new ventures in response to the productivity gains afforded by GenAI. Rather than reducing overall entrepreneurial activity, GenAI appears to \textit{reshape} the landscape—enabling smaller, more agile firms to enter markets once dominated by larger, capital-intensive entities and shifting new firm creation toward leaner organizations, especially in high-AI regions where entrepreneurs are likely to adopt the GenAI tools.
This response among repeat founders underscores the role of GenAI as a ``digital cofounder'' that lowers the managerial and financial thresholds for market entry, thereby compressing the scale distribution of new firms while amplifying the overall rate of business creation.


\section{Placebo and Robustness Tests}
\label{sec:test}

\subsection{Placebo Tests}

While our baseline specification absorbs grid-level and city-by-quarter fixed effects, we implement a set of placebo tests designed to further probe the interpretation of our treatment variable as capturing \textit{AI-specific human capital}—rather than existing local dynamism in innovation and entrepreneurship
in this section.

\subsubsection{AI-Relevant Human Capital vs. General Innovation Capacity}
One hypothesis in our empirical design is that pre-existing AI patent activity proxies for localized \textit{AI-relevant human capital}—that is, the presence of technical talent and domain-specific expertise capable of adopting and productively leveraging GenAI tools. This design underpins our interpretation of the treatment effect: the post-ChatGPT surge in firm entry in high-AI grids arises not merely from general innovation intensity, but from the complementarity between GenAI and pre-existing AI-specific capabilities.

To evaluate this channel more directly, we conduct a placebo test using non-AI patenting activity. The objective is to distinguish the effects of AI-specific knowledge from those driven by broader innovation capacity. If our proposed mechanism is correct, then a similar difference-in-differences design based on non-AI patent strength should yield attenuated or null effects—since general patenting captures innovation capacity not specifically related to AI.

We implement the test in two steps. First, we regress the log number of non-AI patents (filed between 2010 and 2019) on the log number of AI patents at the grid level to remove the shared variation between general and AI-specific innovation activity:
\begin{equation}
\log(\text{nonAI\_patents}_g + 1) = \alpha + \beta \log(\text{AI\_patents}_g + 1) + \varepsilon_g.
\end{equation}
We then classify grids into a \textit{High nonAI} group if their residual $\widehat{\varepsilon}_g$ lies above the 75th percentile—capturing locations with stronger-than-expected non-AI patenting conditional on AI intensity. In other words, these are regions with technological strength \textit{unrelated} to AI. We re-estimate our baseline DiD model with 
\textit{Post} $\times$ \textit{High nonAI} interaction term  instead.

While this placebo test helps isolate the role of AI-specific human capital, we interpret the results conservatively. If our baseline effects are indeed driven by entrepreneurs with AI-relevant expertise—those best positioned to adopt GenAI tools—then the DiD coefficient based on \textit{High nonAI} should shrink substantially relative to the baseline. While we do not expect it to fall exactly to zero—owing to potential measurement error or minor alternative channels—the attenuation itself provides meaningful support for our proposed mechanism. This mirrors the logic of our complementary placebo test using residualized firm counts, where removing the component correlated with AI innovation allows us to better identify the contribution of AI-specific capabilities rather than general entrepreneurial dynamism.

Table~\ref{tab:nonAI} presents the results. Consistent with our hypothesis, 
the coefficient on \textit{Post} $\times$ \textit{High nonAI} is statistically indistinguishable from zero in Columns 1-2, which examine total firm entry and small firm formation. For example, in Column~1, the coefficient is 0.159 (s.e. = 0.478), while in Column 2 it is 0.558 (s.e. = 0.485)—both near zero and not significant. Column 3 shows a marginally significant effect for large firms (–0.427, s.e. = 0.249), but its magnitude is an order of magnitude smaller than the effect observed for AI-intensive grids.

\begin{center}
[Insert \Cref{tab:nonAI} Here]
\end{center}

These findings reinforce the interpretation that the economic impact of GenAI operates primarily through AI-specific human capital rather than general innovation capacity. In particular, the muted response in regions with strong non-AI innovation capacity suggests that it is not general innovation capacity \emph{per se}, but rather the relevance of that capacity to AI, that governs local adoption and entrepreneurial response. This placebo test thus adds credibility to our identification strategy and supports the mechanism of AI-human capital complementarity as the primary channel through which GenAI affects firm formation.

\subsubsection{Residualizing AI Exposure to Address Pre-Existing Entrepreneurial Activity}

Another potential concern is that grid cells with higher AI patent activity may have had higher levels of entrepreneurship prior to the diffusion of GenAI. In that case, our baseline estimates might partially reflect pre-existing local dynamism or unobserved fundamentals, rather than the causal effect of AI-relevant human capital.

To address this concern, we construct a residualization-based placebo test that isolates the component of entrepreneurship uncorrelated with AI-related innovation. Specifically, we regress the logarithm of the average number of new firms in each grid prior to 2019 on the logarithm of the number of AI invention patents filed in 2010–2019:
\begin{equation}
\label{eq:resid_placebo}
\log(\text{FirmEntry}_{g}^{\text{pre}} + 1)
= \alpha + \delta \cdot \log(\text{AI\_Patents}_{g} + 1) + \varepsilon_{g},
\end{equation}
where the residual $\widehat{\varepsilon}_{g}$ captures grid-level entrepreneurial intensity that is linearly uncorrelated with local AI patent activity. We then define a binary indicator, $\textit{HighResid}_{g}$, equal to one for grid cells with above-median residuals, and re-estimate our main difference-in-differences specification, replacing $\textit{HighAI}_{g}$ with $\textit{HighResid}_{g}$.

Table~\ref{tab:residexposure} presents the results. We find that the interaction term $\textit{Post} \times \textit{HighResid}$ remains statistically significant—particularly for small firms (Column~2)—but its magnitude is substantially attenuated relative to the baseline. For example, the coefficient for small-firm entry declines from 7.704 in the main specification in Table~\ref{tab:regression_results} to 1.216 here, representing an approximate 85\% reduction in estimated effect size. Similarly, the coefficient for large-firm entry shrinks in magnitude from $-3.120$ to $-0.481$, also reflecting a roughly 85\% attenuation.

\begin{center}
[Insert \Cref{tab:residexposure} Here]
\end{center}

One thing worth mentioning is that we do not expect this placebo test to yield a null effect. Several factors explain why some residual significance may persist. First, AI patents are a proxy: residual entrepreneurial vibrancy may still embed unmeasured AI-relevant human capital (e.g., in IT services or software startups). Second, the number of pre-2019 firms in a grid—especially in digitally intensive industries—may correlate with latent technological capabilities. Third, residuals from linear regression cannot fully purge nonlinear relationships.

Nevertheless, the attenuation of treatment effects, combined with consistent findings across other placebo and robustness tests, suggests our interpretation that \textit{AI-specific human capital}, not general local vitality, drives the observed post-ChatGPT surge in business formation. This result strengthens the mechanism-based reading of our findings: GenAI facilitates new firm entry most where complementary technical expertise is already in place.

\subsubsection{Random Assignment of AI Exposure Labels}

To further probe the validity of our identification strategy, we implement a placebo test based on random assignment. The goal of this test is to assess whether our main results could be spurious—i.e., whether the observed post-ChatGPT differences in firm formation across high- and low-AI grids are merely artifacts of sample variation, unrelated to actual AI exposure.

Specifically, we randomly assign the \textit{HighAI} label to a set of grids such that the distribution of treated and control observations matches the original sample. We then re-estimate our baseline difference-in-differences specification using these placebo labels. Since this artificial assignment severs any real link between local AI capabilities and GenAI adoption, we would expect no systematic post-period difference in firm entry between the pseudo-treated and control grids.

This placebo test offers a strong falsification check: if our baseline results were driven by chance correlations, model overfitting, or latent spatial trends, we would observe spurious treatment effects even under random assignment. Conversely, the absence of significant coefficients under placebo supports the claim that the original results are driven by meaningful variation in AI-relevant human capital.

\begin{figure}[htbp]
    \centering
    \caption{Random Assignment of AI Exposure Labels}

    \caption*{This figure shows the distribution of estimated coefficients for the interaction term $Post \times HighAI$ across 100 simulations, in which 10,183 grids are randomly assigned to $HighAI=1$. The mean coefficient is 0.064, with a standard deviation of 0.166.} 

    \includegraphics[width=.8\textwidth]{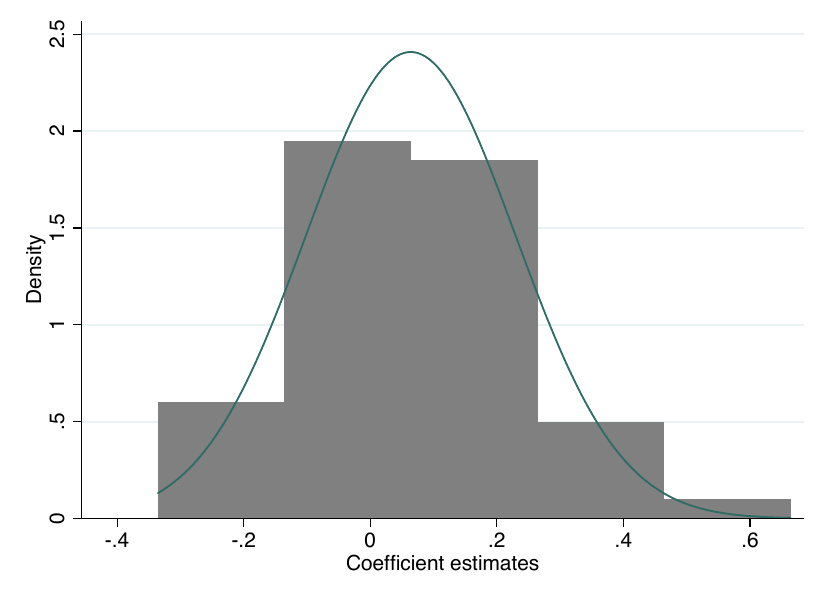}
    
    \label{fig:random_assign}
\end{figure}

We re-estimate our baseline difference-in-differences specification 100 times. Figure~\ref{fig:random_assign} plots the distribution of the resulting $\mathrm{Post}_t \times \mathrm{HighAI}_g$ coefficients. The placebo estimates are tightly centered around zero and statistically insignificant across all outcomes—including total firm entry, small business formation, and large firm entry. The mean coefficient across simulations is 0.064 with a standard deviation of 0.166, indicating no systematic treatment effect under random assignment. This absence of spurious significance reinforces the credibility of our baseline findings and underscores that the observed entrepreneurial response is driven by genuine pre-ChatGPT AI exposure—captured by local AI patenting activity—rather than by chance correlations or unobserved spatial trends.

\begin{center}
[Insert \Cref{fig:random_assign} Here]
\end{center}

\subsection{Additional Robustness Tests}\label{subs:robust_checks}

\subsubsection{Excluding First-Tier Provinces}\label{sec:robust_drop}

A potential concern with our baseline results is that they may be disproportionately driven by a few highly developed regions with concentrated AI activity and vibrant entrepreneurial ecosystems—particularly Beijing, Shanghai, and Guangdong. These provinces account for a large share of China’s innovation output and firm registrations and may exhibit dynamics that are not representative of the broader national landscape.

To address this concern, we re-estimate our difference-in-differences specification after excluding all grid cells located in these three first-tier provinces. Table~\ref{tab:drop} presents the results. Column 1 replicates our baseline specification using the full sample of firm entry as the dependent variable, with city-by-quarter and grid-by-calendar-quarter fixed effects. 
Columns 2-3 examine entry by small and large firms respectively.

Reassuringly, the estimated coefficients on the interaction term \textit{Post} $\times$ \textit{HighAI} remain positive, statistically significant, and similar in magnitude to those in the full-sample analysis. For example, Column 1 reports a coefficient of $4.0$ (s.e. = $1.54$), while Column 2 shows that the effect on small firm formation remains sizable at $6.3$ (s.e. = $1.17$) and significant. The effect on large firms, reported in Column 3, remains negative and significant ($-2.7$, s.e. = $0.76$), consistent with our earlier findings. These results confirm that our main findings are not driven solely by the behavior of a few high-profile innovation hubs. Instead, the observed post-ChatGPT surge in new firm formation—particularly among small entrants in AI-exposed areas—represents a broader pattern that extends beyond China’s most developed provinces. 

\begin{center}
[Insert \Cref{tab:drop} Here]
\end{center}

\subsubsection{Restricting to AI-Active Grids}
\label{sec:robust_aipatent_only}

A natural concern with our baseline difference-in-differences design is that the contrast between high- and low-AI exposure grids may in practice be driven by whether a grid has any AI activity at all.
In the pre-treatment period, a large share of grids report zero AI patenting activity. As a result, our treatment indicator $\mathrm{HighAI}_g$ may conflate intensity of AI exposure with a binary distinction between treated and untreated regions, potentially introducing unobserved differences in baseline entrepreneurship potential.

Table~\ref{tab:a2} presents a robustness check where we restrict the sample to grids that had at least one AI patent prior to the release of ChatGPT. This addresses the concern that our baseline difference-in-differences estimates may conflate the entry response in AI-active areas with structurally different grids that had no prior AI activity. By limiting the sample to AI-active grids, we focus the comparison on areas with varying intensities of prior AI exposure—distinguishing between “high” versus “moderate” AI patenting activity. In this specification, $\mathit{HighAI}_{g}$ is defined as an indicator equal to 1 for grids with more than eight AI patents (the 75th percentile).

The treatment effect remains economically and statistically significant. Column 1 shows that total firm entry in high-AI grids increased by 6.568 firms per grid–quarter relative to moderately exposed grids after the release of ChatGPT (s.e. = 1.962, $p<0.01$). Column 2 reveals that this effect is even larger for small firms, with an estimated coefficient of 11.62 (s.e. = 2.815, $p<0.01$), indicating that the acceleration in firm formation is particularly concentrated among smaller, more agile entrants. In contrast, Column 3 shows a statistically significant and economically meaningful decline in large firm formation, with a coefficient of –5.859 (s.e. = 1.942, $p<0.01$). These effects remain robust across specifications with city-by-quarter and grid-by-calendar-quarter fixed effects.

Overall, this exercise confirms that our main results are not driven by differences between AI-active and AI-inactive areas. Even among locations that had already participated in AI innovation prior to the diffusion of GenAI, those with more intense AI activity experienced significantly stronger boosts in small business formation—and larger declines in large firm entry—following ChatGPT’s release. This pattern reinforces the interpretation that GenAI serves as a complementary force to localized AI knowledge capital, especially in enabling smaller firms to overcome startup frictions and scale new ventures.

Moreover, the dynamic analysis presented earlier (Figure~\ref{fig:dynamic_gpt_firm_entry}) suggests no evidence of pre-treatment divergence between high- and low-AI exposure grids. This lends further support to the identification strategy and mitigates concerns that results are mechanically driven by initial differences in innovation activity or omitted local fundamentals. Overall, this robustness check demonstrates that our core findings are not artifacts of treatment definition or driven by the presence of non-AI grids in the control group.

\subsubsection{Matched Comparison with Nearby Non-AI Grids}

One concern is that the effect might be potentially driven by the national and local government supporting policies in AI as a strategic industry. Throughout 2023, the Chinese government sought to integrate GenAI into broader digital-economy and entrepreneurship policies. Cities such as Beijing, Shanghai, Shenzhen, Hangzhou, and Chengdu introduced subsidy schemes for cloud computing and digital start-ups, promoted “AI + Industry” integration, and established public data-service platforms to support small firms. These initiatives were consistent with China’s long-standing policy orientation toward digital upgrading and innovation-driven growth, but the post-ChatGPT period saw an explicit shift toward the use of AI as an enabling infrastructure for productivity and cost reduction across all sectors rather than as a stand-alone high-tech industry. 

To address this and further assess the robustness of our main finding, we implement a geographically matched comparison design to tackle the systematic differences between AI-active and AI-inactive grids. Specifically, for each grid with at least one AI patent filed between 2010 and 2019, we identify the five geographically nearest grids that had no AI patent activity in the same period, allowing for replacement. This matching strategy ensures a more comparable control group based on spatial proximity, holding constant unobservable regional characteristics that may correlate with economic development, industrial composition, or access to infrastructure.

Table~\ref{tab:a4} presents the difference-in-differences estimates from this matched sample. Column 1 shows that following the release of ChatGPT, new firm entry in AI-active grids increased by 4.345 firms per grid–quarter relative to their spatially matched counterparts (s.e. = 1.354, $p<0.01$). Column 2 demonstrates that this effect is again concentrated among small firms, with an estimated increase of 6.584 firms (s.e. = 1.484, $p<0.01$). In contrast, Column 3 reveals a statistically significant decline in the formation of large firms (coefficient = –2.621, s.e. = 0.693, $p<0.01$), consistent with our previous findings. All specifications include city-by-quarter fixed effects to absorb aggregate shocks, and grid-by-calendar-quarter fixed effects to control for fine-grained temporal and spatial heterogeneity.

\subsubsection{Capital Threshold for Firm Size}
\label{subsec:threshold}
Our baseline definition of small versus large firms relies on a registered capital threshold of one million RMB. While this benchmark is commonly used in administrative classifications, it may not fully capture meaningful differences in firm scale across industries or regions. To ensure that our findings are not mechanically driven by this cutoff, we conduct a robustness check in which we vary the definition of small business across a range of alternative capital thresholds.

Table~\ref{tab:threshold} presents the results using thresholds of 2 million, 3 million, and 5 million RMB. In each case, small firms are defined as those below the specified cutoff, and all specifications include city-by-quarter and grid-by-calendar-quarter fixed effects. Across all three panels, we continue to find strong and statistically significant treatment effects in line with our baseline results. Specifically, the estimated coefficients for \textit{Post $\times$ HighAI} on small firm formation are 7.680 (s.e. = 1.368) at the 2M cutoff, 7.380 (s.e. = 1.384) at 3M, and 7.157 (s.e. = 1.386) at 5M, all significant at the 1\% level. Meanwhile, the corresponding effects for large firms remain significantly negative and decline slightly in magnitude as the threshold increases, ranging from –3.096 (s.e. = 0.473) to –2.573 (s.e. = 0.421).

These patterns confirm that our central conclusion—that GenAI disproportionately stimulates the entry of smaller, more agile firms while displacing or discouraging larger entrants—is robust to alternative classifications of firm size. They also suggest that the observed asymmetry is not an artifact of arbitrary thresholds but rather reflects fundamental differences in how the new AI tools facilitate firms of varying scale.

\section{Conclusion}
\label{sec:conclusion}

This paper provides the first large-scale evidence on how the release of GenAI affects entrepreneurial activity in the real economy. Using comprehensive business registration records for over 12.8 million new firms in China, combined with high-resolution spatial measures of pre-existing AI innovation, we document a substantial increase in new firm formation following the launch of ChatGPT. The effect is concentrated in regions with stronger AI-relevant human capital, driven entirely by small and first-time entrepreneurs. These patterns suggest that GenAI has lowered key barriers to entry, enabling individuals with fewer resources or limited experience to launch new ventures.

Exploring mechanisms, we show that new firms in locations with stronger AI presence become increasingly ``lightweight’’ after the diffusion of GenAI: founders have less prior experience, establish ventures with fewer shareholders, and rely on smaller executive teams. These changes, along with a demonstrated downsizing among repeat founders, suggest that GenAI serves as a substitute for key complementary inputs to entrepreneurship, acting as a ``digital cofounder’’ that enables individuals to launch viable firms with minimal resources. The strongest effects arise in downstream and adoption-oriented sectors—where AI tools can be readily adopted for commercialization and customer-facing activities—rather than in capital- or invention-intensive upstream AI industries.

Taken together, our evidence highlights a new pathway through which general-purpose technologies stimulate economic dynamism—not by expanding frontier innovation, but by democratizing who can start a firm and how ventures are organized at inception. As GenAI tools continue to diffuse and improve, understanding their long-run implications for job creation, entrepreneurial quality, and industrial structure is a central direction for future research. Our findings show that the rise of GenAI represents not only a technological breakthrough, but also an institutional shift in the accessibility, inclusiveness, and scale of entrepreneurship.

\clearpage

\setlength{\bibsep}{0pt plus 0.3ex}
\makeatletter
\patchcmd{\thebibliography}
  {\settowidth}
  {\setlength{\itemsep}{0pt} \setlength{\parskip}{0pt} \settowidth}
  {}{}
\makeatother

\onehalfspacing
\bibliographystyle{apalike}
\bibliography{reference}
\restoregeometry

\section*{Figures \& Tables}
\restoregeometry

\newgeometry{left=1.2cm, right=1.2cm, top=2cm, bottom=2cm}
\restoregeometry

\newgeometry{left=1.2cm, right=1.2cm, top=2cm, bottom=2cm}



\newgeometry{left=1.2cm, right=1.2cm, top=2cm, bottom=2cm}
\restoregeometry

\newgeometry{left=1.2cm, right=1.2cm, top=2cm, bottom=2cm}
\restoregeometry

\newgeometry{left=1.2cm, right=1.2cm, top=2cm, bottom=2cm}
\begin{table}[htbp]
\centering
\caption{Summary Statistics}
\label{tab:summary_statistics}
\caption*{
This table shows the summary statistics of the variables we are interested in at the grid-quarter level. For the definition of the variables, see the Appendix \ref{tab:variable_definition}.
}
\renewcommand{\arraystretch}{1} 
    \begin{tabular}{p{5cm}*{6}{>{\centering\arraybackslash}p{2cm}}}
        \toprule
        \input{main_tables/main_t1}
    \end{tabular}
    
\end{table}
\newpage
\restoregeometry

    
    
\restoregeometry


\newgeometry{left=1.2cm, right=1.2cm, top=2cm, bottom=2cm}
\begin{table}[htbp]
\centering
\caption{Number of New Firms}
\label{tab:regression_results}
\caption*{
This table reports DiD estimates from \Cref{eq:did,eq:did-size}, examining the impact of GenAI on new firm formation. The variable \textit{Post} equals one for quarters after 2022Q4, and \textit{HighAI} is an indicator for grid cells with at least one AI-related patent filed between 2010 and 2019. Columns (1)–(3) use the following dependent variables: (1) number of new firms; (2) number of small firms; and (3) number of large firms. All regressions include city-by-quarter fixed effects and grid-by-calendar-quarter fixed effects to absorb seasonal patterns. Standard errors are clustered at the city level. Variable definitions are provided in Section~\ref{tab:variable_definition}. \textit{Notes:} *** $p<0.01$, ** $p<0.05$, * $p<0.1$.
}
\renewcommand{\arraystretch}{1} 
    \begin{tabular}{p{5cm}*{3}{>{\centering\arraybackslash}p{3cm}}}
        \toprule
        \input{main_tables/main_t2}
    \end{tabular}
\end{table}
\newpage
\restoregeometry

\newgeometry{left=1.2cm, right=1.2cm, top=2cm, bottom=2cm}
\begin{table}[htbp]
\centering
\caption{Robustness: Number of New Firms}
\label{tab:regression_results_winsor}
\caption*{
This table reports DiD estimates from Equation~\eqref{eq:did}, examining the impact of GenAI on new firm formation. The variable \textit{Post} equals one for quarters after 2022Q4, and \textit{HighAI} is an indicator for grid cells with at least one AI-related patent filed between 2010 and 2019. Columns (1)–(6) use the following dependent variables, respectively: (1) and (4) the number of new firms; (2) and (5) the number of small firms; and (3) and (6) the number of large firms. To mitigate the influence of extreme values, we apply a 0.1\% winsorization to the dependent variable. Columns (1)–(3) are winsorized within quarter, while Columns (4)–(6) are winsorized within the full sample. All regressions include city-by-quarter fixed effects and grid-by-calendar-quarter fixed effects to absorb seasonal patterns. Standard errors are clustered at the city level. Variable definitions are provided in  Section~\ref{tab:variable_definition}. \textit{Notes:} *** $p<0.01$, ** $p<0.05$, * $p<0.1$.
}
\renewcommand{\arraystretch}{1} 
    \resizebox{\textwidth}{!}{
    \begin{tabular}{p{4cm}*{6}{>{\centering\arraybackslash}p{2cm}}}
        \toprule
        \input{main_tables/main_t3}
    \end{tabular}
    }
\end{table}
\newpage
\restoregeometry

\newgeometry{left=1.2cm, right=1.2cm, top=2cm, bottom=2cm}
\begin{table}[htbp]
\centering
\caption{Industry-Specific DID Estimates}
\label{tab:industry-heterogeneity}
\caption*{
This table presents the estimated coefficients of the interaction term from Equation~\eqref{eq:did}, assessing the impact of GenAI on new firm formation across 96 industries. Industries are sorted in descending order by the estimated coefficient. Panel A presents the top 15 industries, and Panel B presents the bottom 15 industries.
}
\renewcommand{\arraystretch}{1} 
\textbf{Panel A: Top 15} \\[0.3em]
    \begin{tabular}{p{8cm}*{4}{>{\centering\arraybackslash}p{2cm}}}
        \toprule
        \input{main_tables/main_t5}
    \end{tabular}


\renewcommand{\arraystretch}{1} 
\textbf{Panel B: Bottom 15} \\[0.3em]
    \begin{tabular}{p{8cm}*{4}{>{\centering\arraybackslash}p{2cm}}}
        \toprule
        \input{main_tables/main_t5_bottom}
    \end{tabular}     
\end{table}
\newpage
\restoregeometry

\newgeometry{left=1.2cm, right=1.2cm, top=2cm, bottom=2cm}
\begin{table}[htbp]
\centering
\caption{Industry Heterogeneity: Relevance to AI}
\label{tab:industry}
\caption*{
This table reports estimates of changes in the number of new firms across different industries following the release of ChatGPT. \textit{Post} is an indicator equal to one for quarters after 2022Q4, and \textit{HighAI} is a dummy equal to one for grids with at least one AI patent filed between 2010 and 2019. Columns (1)–(6) use the following dependent variables, respectively: (1) number of new firms in industries more likely to be upstream of AI; (2) number of new firms in industries less likely to be upstream of AI; (3) number of new firms in industries more likely to be downstream of AI; (4) number of new firms in industries less likely to be downstream of AI; (5) number of new firms in industries with high entrepreneurship scores; and (6) number of new firms in industries with low entrepreneurship scores. All specifications include city-by-quarter fixed effects and grid-by-calendar-quarter fixed effects. Standard errors are clustered at the city level. Variable definitions are provided in  Section~\ref{tab:variable_definition}. \textit{Notes:} *** $p<0.01$, ** $p<0.05$, * $p<0.1$.
}
\renewcommand{\arraystretch}{1} 
    \resizebox{\textwidth}{!}{
    \begin{tabular}{p{4cm}*{8}{>{\centering\arraybackslash}p{2cm}}}
        \toprule
        \input{main_tables/main_t4}
    \end{tabular}
    }
\end{table}
\newpage
\restoregeometry

\newgeometry{left=1.2cm, right=1.2cm, top=2cm, bottom=2cm}
\begin{table}[htbp]
\centering
\caption{ChatGPT and Serial Entrepreneurship}
\label{tab:serial_entrepreneurship}
\caption*{
This table reports estimates of changes in the ratio of serial entrepreneurs following the release of ChatGPT. \textit{Post} is an indicator equal to one for quarters after 2022Q4, and \textit{HighAI} is a dummy equal to one for grids with at least one AI patent filed between 2010 and 2019. The ratio of serial entrepreneurs is defined as the percentage of new firm founders (legal representatives) who had established at least one other firm within the three years preceding the current firm’s founding, excluding the month of establishment. Columns (1)–(3) use the following dependent variables, respectively: (1) the ratio of serial entrepreneurs in the full sample; (2) the ratio among small firms; and (3) the ratio among large firms. All specifications include city-by-quarter fixed effects and grid-by-calendar-quarter fixed effects. Standard errors are clustered at the city level. Variable definitions are provided in  Section~\ref{tab:variable_definition}. \textit{Notes:} *** $p<0.01$, ** $p<0.05$, * $p<0.1$.
}
\renewcommand{\arraystretch}{1} 
    \begin{tabular}{p{4cm}*{4}{>{\centering\arraybackslash}p{3cm}}}
        \toprule
        \input{main_tables/main_t6}
    \end{tabular}
\end{table}
\newpage
\restoregeometry

\newgeometry{left=1.2cm, right=1.2cm, top=2cm, bottom=2cm}
\begin{table}[htbp]
\centering
\caption{ChatGPT and Shareholder Numbers}
\label{tab:shareholder}
\caption*{
This table reports estimates of changes in the number of shareholders following the release of ChatGPT. \textit{Post} is an indicator equal to one for quarters after 2022Q4, and \textit{HighAI} is a dummy equal to one for grids with at least one AI patent filed between 2010 and 2019. Panel A reports results for the total number of shareholders (including both individual and corporate shareholders), while Panel B focuses on the share of individual shareholders. Column (1) presents results for the full sample, Column (2) for small firms, and Column (3) for large firms. All specifications include city-by-quarter fixed effects and grid-by-calendar-quarter fixed effects. Standard errors are clustered at the city level. Variable definitions are provided in  Section~\ref{tab:variable_definition}. \textit{Notes:} *** $p<0.01$, ** $p<0.05$, * $p<0.1$.
}
\renewcommand{\arraystretch}{1} 
    \textbf{Panel A: Total Shareholders} \\[0.3em]
    \begin{tabular}{p{4cm}*{4}{>{\centering\arraybackslash}p{3cm}}}
        \toprule
        \input{main_tables/main_t7_panelA}
    \end{tabular}
    
    \vspace{1em} 
    
    \textbf{Panel B: Percentage of Individual Shareholders} \\[0.3em]
    \begin{tabular}{p{4cm}*{4}{>{\centering\arraybackslash}p{3cm}}}
        \toprule
        \input{main_tables/main_t7_panelB}
    \end{tabular}
\end{table}
\newpage
\restoregeometry

\newgeometry{left=1.2cm, right=1.2cm, top=2cm, bottom=2cm}
\begin{table}[htbp]
\centering
\caption{ChatGPT and Executive Member}
\label{tab:executive}
\caption*{
This table reports estimates of changes in the number of executive members following the release of ChatGPT. \textit{Post} is an indicator equal to one for quarters after 2022Q4, and \textit{HighAI} is a dummy equal to one for grids with at least one AI patent filed between 2010 and 2019. Columns (1)–(3) use the following dependent variables, respectively: (1) number of executive members; (2) number of executive members in small firms; and (3) number of executive members in large firms. All specifications include city-by-quarter fixed effects and grid-by-calendar-quarter fixed effects. Standard errors are clustered at the city level. Variable definitions are provided in  Section~\ref{tab:variable_definition}. \textit{Notes:} *** $p<0.01$, ** $p<0.05$, * $p<0.1$.
}
\renewcommand{\arraystretch}{1} 
    \begin{tabular}{p{4cm}*{4}{>{\centering\arraybackslash}p{3cm}}}
        \toprule
        \input{main_tables/main_t8}
    \end{tabular}
\end{table}
\newpage
\restoregeometry

\newgeometry{left=1.2cm, right=1.2cm, top=2cm, bottom=2cm}
\begin{table}[htbp]
\centering
\caption{Serial Entrepreneurship and Substitution Towards Smaller Firms}
\label{tab:size_shrinkage_for_cumbuersom}
\caption*{
This table reports estimates of changes in the relative size of new firms established by serial entrepreneurs (legal representatives who had founded at least one other firm within the three years preceding the current firm’s establishment, excluding the month of founding). \textit{Post} is an indicator equal to one for quarters after 2022Q4, and \textit{HighAI} is a dummy equal to one for grids with at least one AI patent filed between 2010 and 2019. Panel A reports the grid-quarter level percentage of new firms established by a serial entrepreneur whose registered capital exceeds that of the entrepreneur’s previous firm. Panel B reports the grid-quarter average ratio of registered capital in the new firm established by a serial entrepreneur to that in the previous firm. To mitigate the impact of outliers, the top 1\% and bottom 1\% of values are trimmed. Columns (1) presents the result for all firms, Column (2) for small firms, and Column (3) for large firms. All specifications include city-by-quarter fixed effects and grid-by-calendar-quarter fixed effects. Standard errors are clustered at the city level. Variable definitions are provided in  Section~\ref{tab:variable_definition}. \textit{Notes:} *** $p<0.01$, ** $p<0.05$, * $p<0.1$.
}
\renewcommand{\arraystretch}{1} 
    \textbf{Panel A: Percentage of New Firms Established by Serial Entrepreneurs with Larger Registered Capital} \\[0.3em]
    \begin{tabular}{p{4cm}*{4}{>{\centering\arraybackslash}p{3cm}}}
        \toprule
        \input{appendix_tables/appendix_a5_panelA}
    \end{tabular}
    
    \vspace{1em} 
    
    \textbf{Panel B: Average Capital Ratio of New to Prior Firms for Serial Entrepreneurs} \\[0.3em]
    \begin{tabular}{p{4cm}*{4}{>{\centering\arraybackslash}p{3cm}}}
        \toprule
        \input{appendix_tables/appendix_a5_panelB}
    \end{tabular}
\end{table}
\newpage
\restoregeometry

\newgeometry{left=1.2cm, right=1.2cm, top=2cm, bottom=2cm}
\begin{table}[htbp]
\centering
\caption{Placebo Tests Using NonAI Patents}
\label{tab:nonAI}
\caption*{
This table reports estimates of changes in the number of new firms following the release of ChatGPT. \textit{Post} is an indicator equal to one for quarters after 2022Q4.
In this table, we focus on grids with non-AI patents and construct \textit{High nonAI} based on the residuals from regressing the log number of non-AI patents on the log number of AI patents. \textit{High nonAI} equals one for grids whose residual exceeds the 75th percentile. Columns (1)–(3) use the following dependent variables: (1) number of new firms; (2) number of small firms; and (3) number of large firms. All regressions include city-by-quarter fixed effects and grid-by-calendar-quarter fixed effects to absorb seasonal patterns. Standard errors are clustered at the city level. Variable definitions are provided in  Section~\ref{tab:variable_definition}. \textit{Notes:} *** $p<0.01$, ** $p<0.05$, * $p<0.1$.
}
\renewcommand{\arraystretch}{1} 
    \begin{tabular}{p{4cm}*{3}{>{\centering\arraybackslash}p{3cm}}}
        \toprule
        \input{main_tables/main_t10}
    \end{tabular}
\end{table}
\newpage
\restoregeometry

\newgeometry{left=1.2cm, right=1.2cm, top=2cm, bottom=2cm}
\begin{table}[htbp]
\centering
\caption{Residualizing AI Exposure to Remove Entrepreneurial Correlation}
\label{tab:residexposure}
\caption*{
This table reports estimates of changes in the number of new firms following the release of ChatGPT. \textit{Post} is an indicator equal to one for quarters after 2022Q4. \textit{HighResid} is constructed from the residuals of a regression of the log number of prior firms (plus one) on the log number of AI patents (plus one). \textit{HighResid} equals one for grid cells with residuals above the median. Columns (1)–(3) use the following dependent variables: (1) number of new firms; (2) number of small firms; and (3) number of large firms. All regressions include city-by-quarter fixed effects and grid-by-calendar-quarter fixed effects to absorb seasonal patterns. Standard errors are clustered at the city level. Variable definitions are provided in  Section~\ref{tab:variable_definition}. \textit{Notes:} *** $p<0.01$, ** $p<0.05$, * $p<0.1$.
}
\renewcommand{\arraystretch}{1} 
    \begin{tabular}{p{4cm}*{3}{>{\centering\arraybackslash}p{3cm}}}
        \toprule
        \input{main_tables/main_t11}
    \end{tabular}
\end{table}
\newpage
\restoregeometry

\newgeometry{left=1.2cm, right=1.2cm, top=2cm, bottom=2cm}
\begin{table}[htbp]
\centering
\caption{Excluding First-Tier Provinces}
\label{tab:drop}
\caption*{
This table reports estimates of changes in the number of new firms following the release of ChatGPT. \textit{Post} is an indicator equal to one for quarters after 2022Q4, and \textit{HighAI} is a dummy equal to one for grids with at least one AI patent filed between 2010 and 2019. To mitigate the influence of highly active regions, we exclude grids located in Beijing, Shanghai, and Guangdong. Columns (1)–(3) use the following dependent variables: (1) number of new firms; (2) number of small firms; and (3) number of large firms. All regressions include city-by-quarter fixed effects and grid-by-calendar-quarter fixed effects to absorb seasonal patterns. Standard errors are clustered at the city level. Variable definitions are provided in  Section~\ref{tab:variable_definition}. \textit{Notes:} *** $p<0.01$, ** $p<0.05$, * $p<0.1$.
}
\renewcommand{\arraystretch}{1} 
    \begin{tabular}{p{5cm}*{3}{>{\centering\arraybackslash}p{3cm}}}
        \toprule
        \input{main_tables/main_t9}
    \end{tabular}
\end{table}
\newpage
\restoregeometry

\appendix
\counterwithin{figure}{section}

\newgeometry{left=1.2cm, right=1.2cm, top=2cm, bottom=2cm}
\section{Appendix A}
\label{sec:appendix-a}
\subsection{Variables and Definitions} \label{tab:variable_definition}

\begin{table}[h]
\centering
\renewcommand{\arraystretch}{1} 
\resizebox{.91\textwidth}{!}{
    \begin{tabular}{p{4.8cm}*{1}{>{\arraybackslash}p{13.2cm}}}
        \toprule
        \input{main_tables/variable_def}
    \end{tabular}
    }
\end{table}
\newpage
\restoregeometry

\subsection{AI relevance Score}\label{sec:AI-relevance-score}

This section outlines the construction and validation of the firm-level AI relevance scores that quantify (i) how tightly a firm’s business activities connect to AI technologies and applications (\emph{upstream} and \emph{downstream}) and (ii) the extent to which AI can enable \emph{entrepreneurship}. 
The construction proceeds in two steps. 
First, we represent each firm’s business scope as a combination of latent \emph{textual factors}, i.e., topics extracted from firm's business description that capture the semantic structure and economic meaning of its activities (see \Cref{sub:business_topics}). In the second step, we employ the state-of-the-art large language model GPT-4o \citep{hurst2024gpt} to assign topic-specific AI relevance scores.
A firm’s AI relevance scores are then computed as a weighted average of these topic scores, with weights given by the firm’s topic loadings (\Cref{sec:chatgpt_annotation}).

We check the validity of the AI relevance scores in \Cref{sub:validation}. We randomly sample a subset of firms, review their business websites and publicly available information, and compare these qualitative assessments with our constructed AI-relevance scores. 
Importantly, we do not reuse the business-scope text in construction, thereby avoiding circularity.
The clear correspondence between the two provides supporting evidence for validity of our constructed AI relevance measures.

\subsubsection{Business Topics}\label{sub:business_topics}
Let \( N \) denote the set of all firms that entered during the sample period, and let \( N_t \) represent the set of entrants in quarter \( t \). The full sample of entrants can therefore be expressed as the union across all quarters:
\[
N = \bigcup_t N_t.
\]
We denote \(|N|\) as the total number of new firms in the sample and \(|N_t|\) as the number of entrants in quarter \( t \).

Each firm in our dataset is associated with a paragraph describing its business scope, which we treat as a short textual document. Let \( V \) denote the set of all unique vocabulary terms appearing across these documents after standard text preprocessing (e.g., removal of stop words), and let \(|V|\) be its cardinality. To extract structured information from this corpus, we adopt the topic modeling framework developed by \citet{cong2025textual}, which represents each document as a mixture of latent topics and each topic as a distribution over vocabulary terms. This approach enhances the interpretability of traditional topic models while maintaining scalability for large corpora. The construction proceeds in three steps:

\begin{itemize}
    \item \textbf{Word Embedding.} We encode the semantic meaning of each term using the pretrained Chinese embedding model \texttt{bge-base-zh-v1.5}, developed by the Beijing Academy of Artificial Intelligence \citep{xiao2024c}. This model maps each word into a dense vector space that captures both syntactic and semantic relationships.

    \item \textbf{Term Clustering.} We then apply Locality-Sensitive Hashing (LSH) to cluster these word vectors into \( K \) mutually exclusive groups, each containing at most 50 terms. Every word is assigned to exactly one cluster. This clustering step groups semantically similar words—based on their embedding proximity—into coherent economic concepts, improving the interpretability of the resulting topics.

    \item \textbf{Topic Extraction.} For each cluster \( k \in \{1, \dots, K\} \), let \( V_k \subset V \) denote the subset of vocabulary terms assigned to cluster \( k \). We construct a document-term matrix using only the terms in \( V_k \) and apply Latent Semantic Analysis (LSA) via singular value decomposition (SVD). This yields topic loadings for each document, denoted \( \theta_{i,k} \) for firm \( i \), and word loadings for each topic, denoted \( \bm{f}_k \).
\end{itemize}

Before implementing these steps, we clean all business scope descriptions by removing boilerplate disclaimers and tokenize the text using Jieba. We then cluster the unique terms, producing an initial set of 8,460 clusters. To enhance computational efficiency, we estimate topics using a 5\% random sample of firms established between 2015 and 2024, comprising 3,019,725 documents. Clusters with fewer than five terms are removed, leaving 4,967 economically interpretable topics. Finally, we compute topic loadings for all firms in the universe following \citet{cong2025textual}. For each firm \( i \), we calculate a normalized word frequency vector and multiply it by the topic-specific term distribution, yielding the topic loading \( \theta_{i,k} \), which represents the firm’s loading in topic \( k \).

\subsubsection{AI Relevance via Large Language Models}
\label{sec:chatgpt_annotation}
This subsection describes how we generate a structured and semantically consistent evaluation of each topic’s relationship to artificial intelligence using large language models (LLMs). Specifically, we employ GPT-4o \citep{hurst2024gpt} to assess each topic along multiple dimensions of AI relevance, based on the topic’s most representative keywords and their associated weights derived from the topic model.

For each topic, we construct a concise text prompt containing a ranked list of top keywords and their corresponding importance weights. We then instruct the model to evaluate four conceptual dimensions, each scored on a scale from $-100$ to $100$:

\begin{enumerate}
    \item \textbf{AI Upstream Score}: the extent to which the topic concerns upstream AI activities—such as model development, data infrastructure, computing resources, or foundational technologies.
    \item \textbf{AI Downstream Score}: the extent to which the topic captures downstream AI activities—such as application development, product integration, or end-user services.
    \item \textbf{Entrepreneurship Helpfulness Score}: the degree to which the topic is relevant or enabling for entrepreneurial or solo creative activities (e.g., content creation, personal branding, low-cost experimentation, or digital marketing), scored as:
    \begin{itemize}
        \item $-100 =$ obstructive or irrelevant to entrepreneurship,
        \item $0 =$ neutral or unrelated,
        \item $100 =$ highly enabling for entrepreneurial or small-scale creative work.
    \end{itemize}
\end{enumerate}

For each topic, the model also provides a brief justification and, when appropriate, examples of firms or industries where the topic is particularly relevant to AI-driven entrepreneurship. These responses are parsed and compiled into a structured dataset containing the three topic-level scores—\emph{AI Upstream}, \emph{AI Downstream}, and \emph{Entrepreneurship Helpfulness}. To ensure robustness, the evaluation can be repeated multiple times to assess the stability of the scores.

Finally, we aggregate the topic-level evaluations into firm-level measures. For firm \( i \), the AI-relevance score is computed as a weighted average of the topic-specific scores:
\[
\text{AIscore}_{i} = \frac{\sum_{k} \theta_{i,k} s_k}{\sum_{k} \theta_{i,k}},
\]
where \( \theta_{i,k} \) denotes the topic loading for firm \( i \) on topic \( k \), and \( s_k \) is the score assigned to topic \( k \) by GPT-4o. This approach yields a continuous, interpretable measure of each firm’s exposure to AI-related activities and its potential entrepreneurial complementarity with GenAI technologies.

\subsubsection{Validation of the AI Relevance Score}\label{sub:validation}

This subsection validates the AI relevance score constructed above. Since LLM pipelines can suffer from hallucination and bias, we design a two-stage, human-in-the-loop procedure that leverages firms' webpages with deterministic preprocessing, conservative fallbacks, and manual audits. 
Importantly, we do \emph{not} reuse the ``business scope'' descriptions that are used to construct the AI relevance score so as to provide a clean validation. 
We begin from a random sample of 27,239 firms and validate the scoring by comparing the keywords between the high- and low-score groups.

\paragraph{A Two-stage Procedure.}

We design a two-stage procedure for the validation: in Stage 1, we crawl firms' webpages and extract raw keywords; and in Stage 2, we construct high-level keywords.

\begin{enumerate}
    \item \textbf{Stage 1: Websites collection and keywords extraction}
    
We first obtain each firm’s official website and crawl its site for the initial random  samples of 27,239 firms. During preprocessing, we remove malformed entries and uninformative URLs, such as using a generic e-commerce domain as the company homepage (e.g., \texttt{www.taobao.com}) while allowing store-specific URLs on such platforms. We then perform domain normalization and correction, retaining 24,528 firms. 

We then identify and fetch the page most suitable for a business description by combining \texttt{sitemap.xml} cues, common “About/Business Introduction” paths (such as \texttt{/about}), homepage link analysis, and a weighted scoring scheme, with the homepage as a fallback.
If the firm’s site is itself a path, as is common for e-commerce storefronts, we use it directly. 
We tried both simple HTTP requests and a headless browser (\texttt{Playwright}), both of which perform similarly.
This step yields 8,729 useful websites.

To extract raw keywords, from each selected page, we convert HTML to clean text using \texttt{trafilatura} and then transform it into structured fields via batched LLM prompts that summarize the firm’s business description, primary industry, and finally extract raw keywords. We obtain 8,871 raw keywords from all the firms.

\item \textbf{Stage 2: Human-audited semantic clustering and keyword taxonomy}

In Stage 2, we consolidate the 8,871 raw keywords to construct a concise, high-level taxonomy, yielding 129 high-level keywords that are granular yet broad enough.

Specifically, we embed keywords with a multi-lingual LLM model \texttt{bge-m3} \citep{chen2024bge}, reduce dimensionality using UMAP \citep{mcinnes2018umap}, and form clusters with HDBSCAN \citep{mcinnes2017hdbscan}. 
To summarize the cluster, we involve human in the loop: we adopt a semi-automated, semi-manual workflow where AI proposes concise, cluster-level descriptors and human reviewers audit, refine, and, where necessary, conduct secondary clustering and re-summarization for mixed or oversized clusters. The result is a validated set of 129 high-level keywords that are granular yet broadly interpretable. For example, “data science,” “Internet-of-Things,” and ``real estate.''

\end{enumerate}

\paragraph{Results.}

To assess whether the AI relevance score is sensible, we compute for each keyword its relative frequency in the high- versus low-score groups; equivalently, we examine the ratio of keyword frequency in the high group to that in the low group.
We highlight the top 20 with the largest ratios as the most differentiating terms by the three AI relevance scores respectively.

\begin{enumerate}
    \item \textbf{AI upstream score.} 

Figure \ref{fig:upstream-freq} shows the relative frequency of keywords in firms’ website descriptions for the high- versus low–AI-upstream groups. In the high-upstream panel (Figure \ref{fig:high-upstream-freq}), the top keywords concentrate on the upstream industries of AI: data industry, communications networks, and the Internet of Things, followed by science- and hardware-intensive enablers such as biotechnology/drug R\&D, intelligent manufacturing, consumer/electronic components, aviation/medical diagnostics, new energy, and semiconductors. These categories map naturally to data generation and curation, connectivity, sensing, and compute-adjacent hardware, which are precisely the upstream layers of the AI stack. 

By contrast, the low-upstream panel (Figure \ref{fig:low-upstream-freq}) is dominated by traditional, service-oriented, or non-digital verticals: food industry, logistics, cultural industry, landscaping, customer service, leasing, interior design, apparel, and lodging, which are are largely irrelevant to AI inputs or plausibly downstream users of technology rather than suppliers of core AI inputs.

Taken together, these keywords distributions based on the their websites contrast the upstream enablers in the high-scored group with the conventional brick-and-mortar activities in the low-scored group, providing evidence that the AI upstream score captures the intended construct.

    \item \textbf{AI downstream score.} 

Figure \ref{fig:downstream-freq} plots the relative frequency of keywords in firms’ website descriptions for the high- versus low–AI downstream groups. In the high-downstream panel (Figure \ref{fig:high-downstream-freq}), top terms concentrate on end-user applications and distribution channels, including online games, office collaboration, media/entertainment, dating/relationships, image processing, information services, advertising/marketing—and on consumer-facing electronics. These are demand-side domains where AI can be directly embedded in products and user experiences.

By contrast, the low-downstream panel (Figure \ref{fig:low-downstream-freq}) is dominated by traditional, non-digital sectors, such as leasing, construction, food, interior design/renovation, cultural industry, landscaping, apparel, lodging, which have weaker direct ties to AI products. These patterns support the validity of the downstream score.

    \item \textbf{Entrepreneurship score.} 

Figure \ref{fig:ent-freq} reports the analogous analysis for the entrepreneurship score. In the high-entrepreneur panel (Figure \ref{fig:high-ent-freq}), top keywords cover internet markets and tools: computation power, consumer electronics, online games, media/entertainment, marketing, website/information services, translation services, artificial intelligence, ``no-code'' platforms (services that allows users to create applications, websites, and automations through visual interfaces and drag-and-drop tools instead of writing traditional programming code), information security, and general internet/O2O. These categories characterize rapid software iteration, low distribution costs, and scalable monetization, which are conducive to startup. 

The low-entrepreneur panel (Figure \ref{fig:low-ent-freq}) instead features asset-intensive industries with long capital cycles: metals, civil engineering, construction, chemical manufacturing, recycling, materials, logistics, supply chain, import/export, warehousing, project quoting, electric power, real estate, shipping, and mining, which are less conducive to early-stage, software-centric entrepreneurship. This contrast shows that the entrepreneur score also captures the intended construct.

\begin{figure}[h]
\centering
\begin{subfigure}{.5\textwidth}
  \centering
  \includegraphics[width=\linewidth]{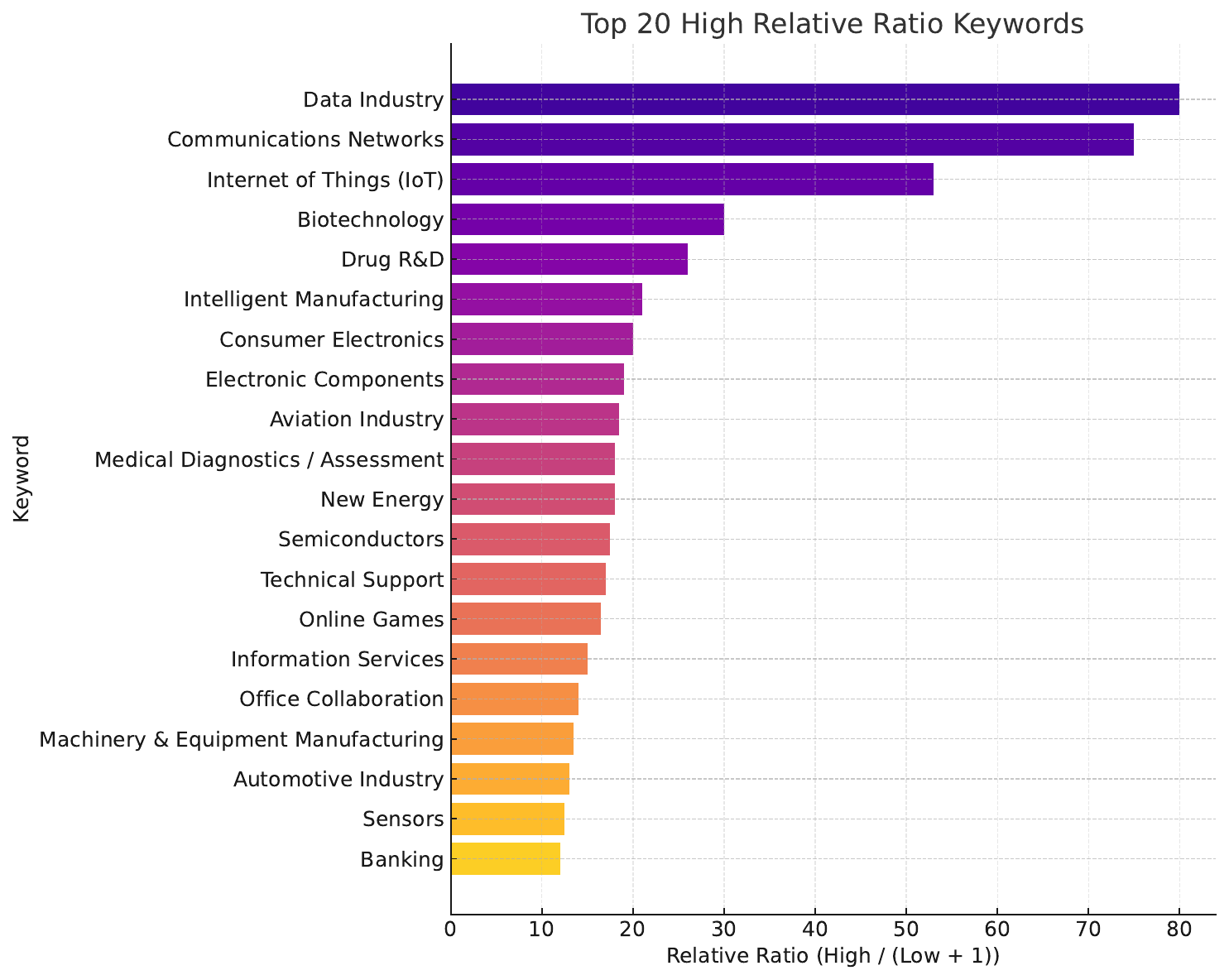}
  \caption{High AI upstream.}
  \label{fig:high-upstream-freq}
\end{subfigure}%
\begin{subfigure}{.5\textwidth}
  \centering
  \includegraphics[width=\linewidth]{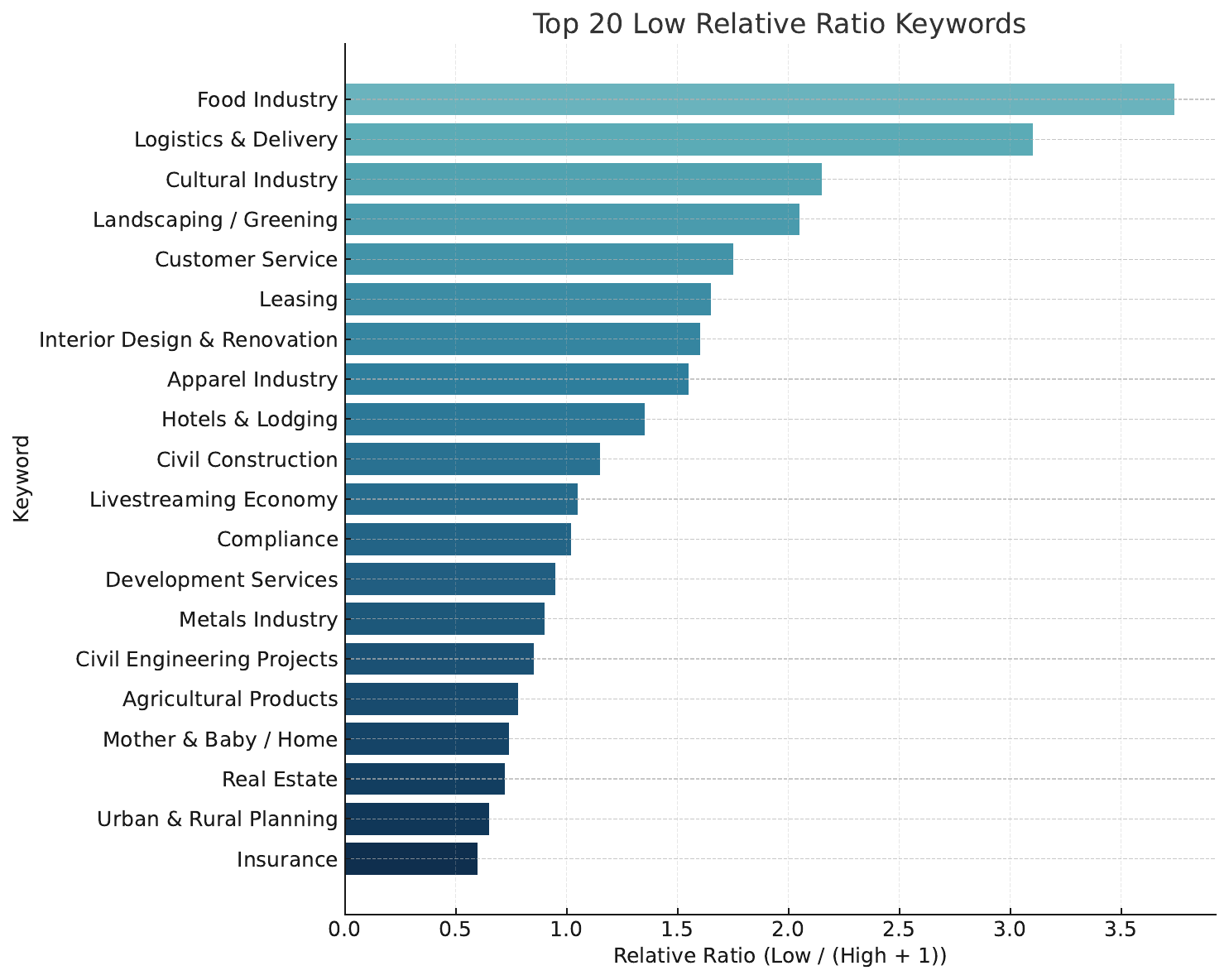}
  \caption{Low AI upstream.}
  \label{fig:low-upstream-freq}
\end{subfigure}
\caption{Relative frequency in the high- versus low- AI upstream groups.}
\label{fig:upstream-freq}
\end{figure}

\begin{figure}[h]
\centering
\begin{subfigure}{.5\textwidth}
  \centering
  \includegraphics[width=\linewidth]{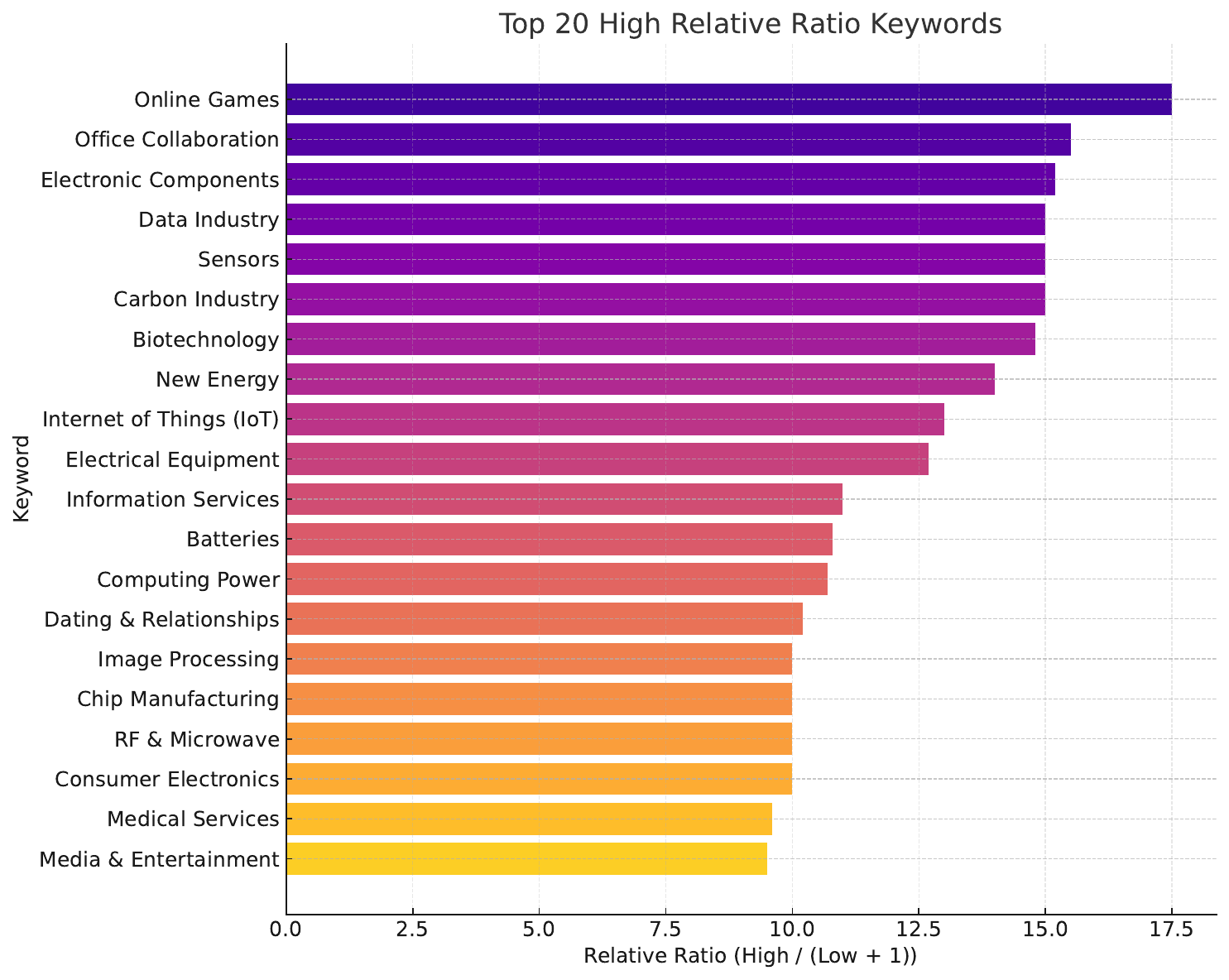}
  \caption{High AI downstream.}
  \label{fig:high-downstream-freq}
\end{subfigure}%
\begin{subfigure}{.5\textwidth}
  \centering
  \includegraphics[width=\linewidth]{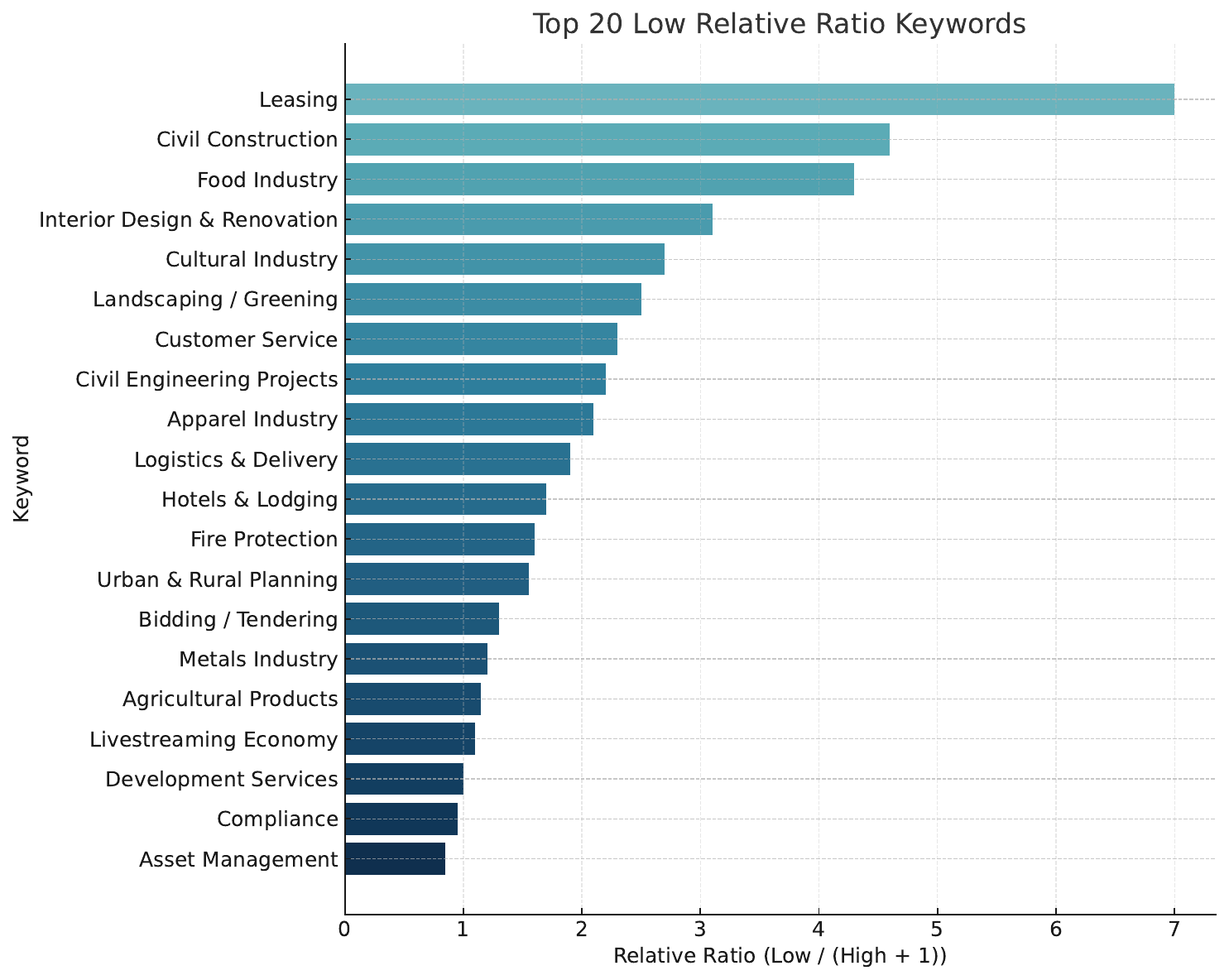}
  \caption{Low AI downstream.}
  \label{fig:low-downstream-freq}
\end{subfigure}
\caption{Relative frequency in the high- versus low- AI downstream groups.}
\label{fig:downstream-freq}
\end{figure}

\begin{figure}[h]
\centering
\begin{subfigure}{.5\textwidth}
  \centering
  \includegraphics[width=\linewidth]{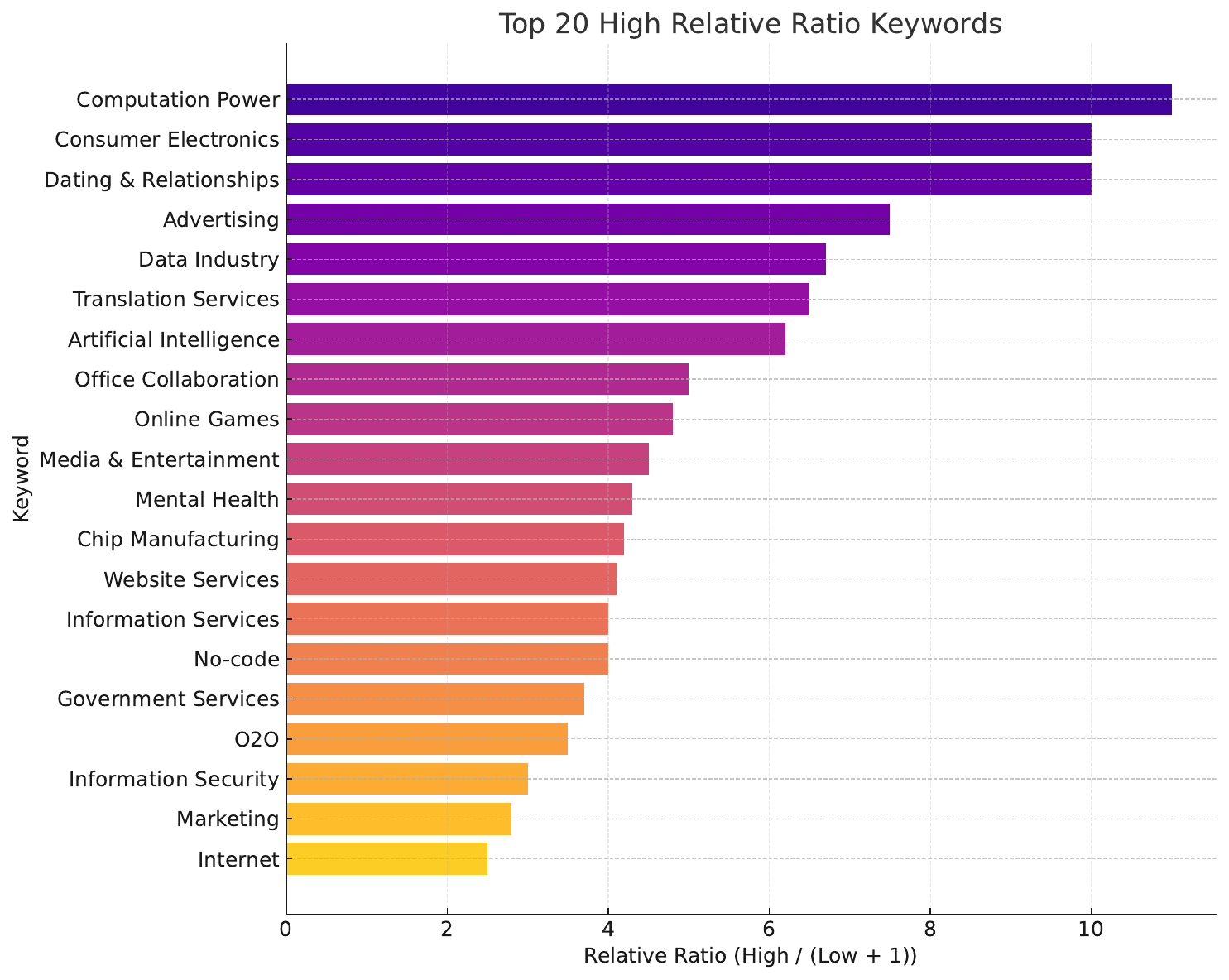}
  \caption{High AI-entrepreneurship.}
  \label{fig:high-ent-freq}
\end{subfigure}%
\begin{subfigure}{.5\textwidth}
  \centering
  \includegraphics[width=\linewidth]{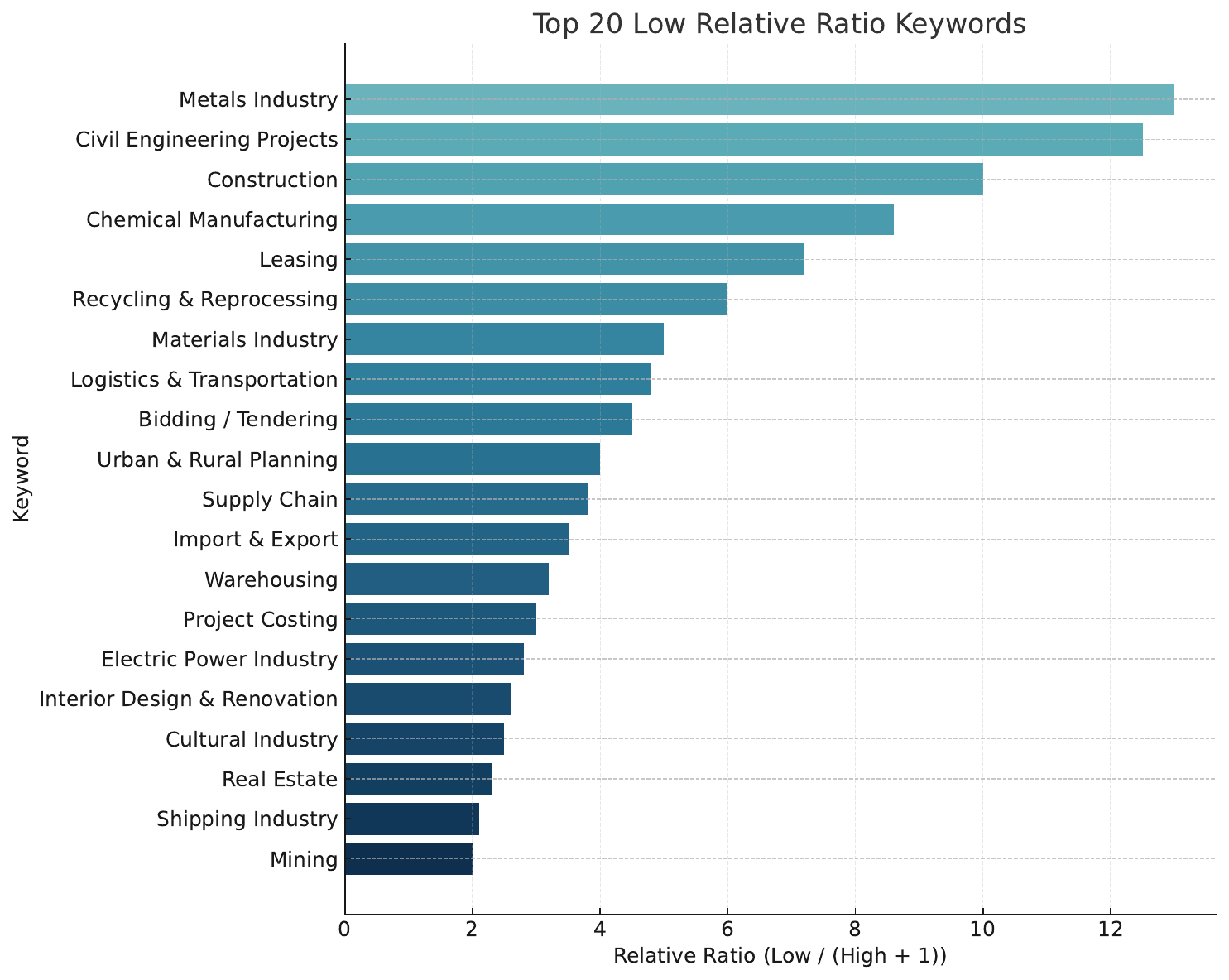}
  \caption{Low AI-entrepreneurship.}
  \label{fig:low-ent-freq}
\end{subfigure}
\caption{Relative frequency in the high- versus low- AI entrepreneurship groups.}
\label{fig:ent-freq}
\end{figure}

\end{enumerate}

\newgeometry{left=1.2cm, right=1.2cm, top=2cm, bottom=2cm}
\section{Appendix B}

\newgeometry{left=1.2cm, right=1.2cm, top=2cm, bottom=2cm}
\begin{figure}[htbp]
    \centering
    \caption{Geographical Distribution of AI Patents in Major Cities}
    \label{fig:beijing-shanghai-shenzhen}
    \vspace{0.5em}
    \caption*{
    This figure plots the spatial distribution of AI patents across three representative cities—Beijing, Shanghai, and Shenzhen—using H3 grid cells at 2km resolution. Each point represents a hexagonal cell, with shading intensity power-normalized to reflect local AI patent density during the pre-ChatGPT period (2010–2019). The maps highlight substantial within-city heterogeneity in AI innovation activity, even within major metropolitan areas.
    }

    \vspace{1em}

    \begin{subfigure}[t]{0.45\textwidth}
        \centering
        \includegraphics[width=\textwidth]{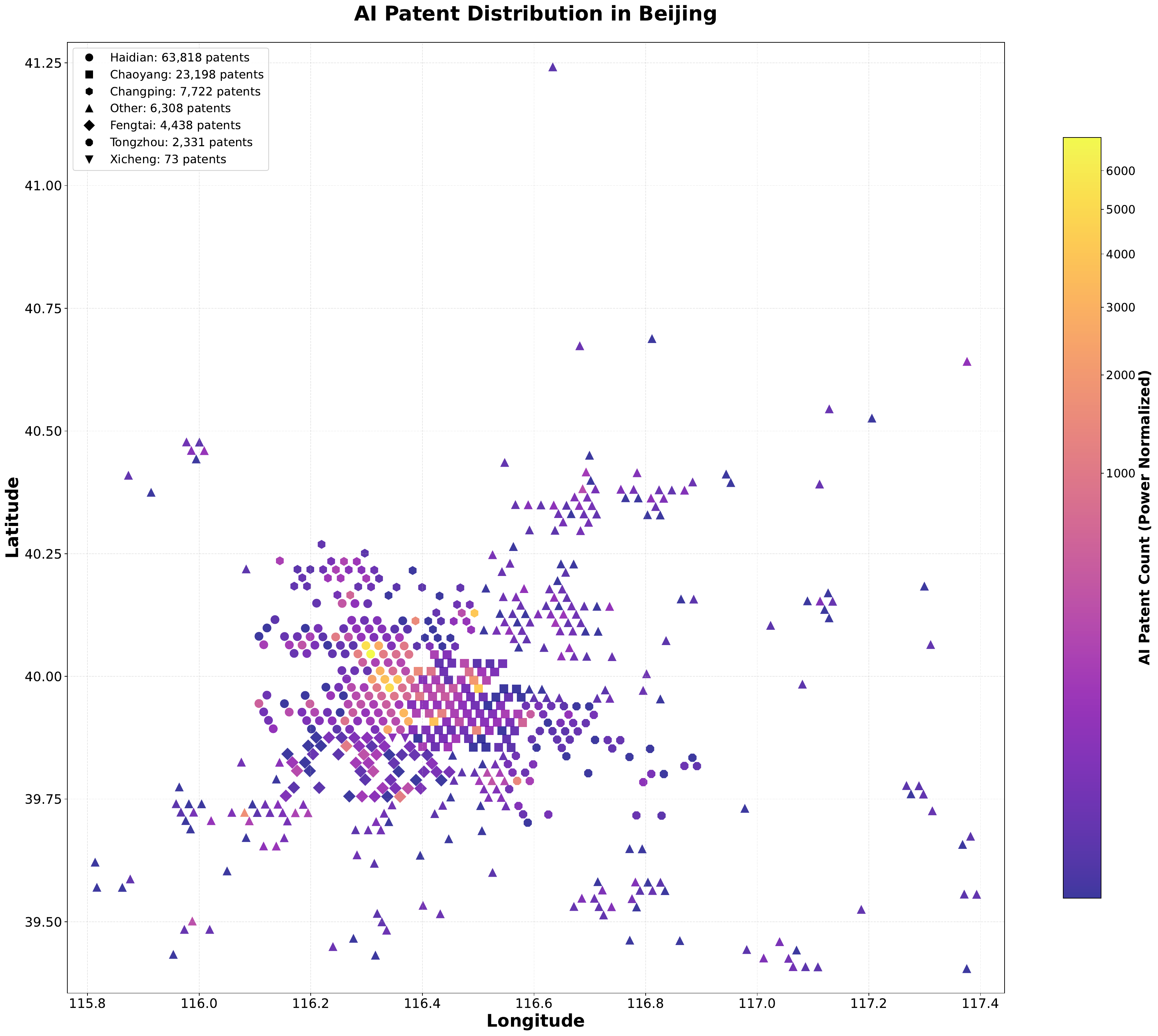}
        \caption{Panel A: Beijing}
        \label{fig:panel_beijing}
    \end{subfigure}
    \hfill
    \begin{subfigure}[t]{0.45\textwidth}
        \centering
        \includegraphics[width=\textwidth]{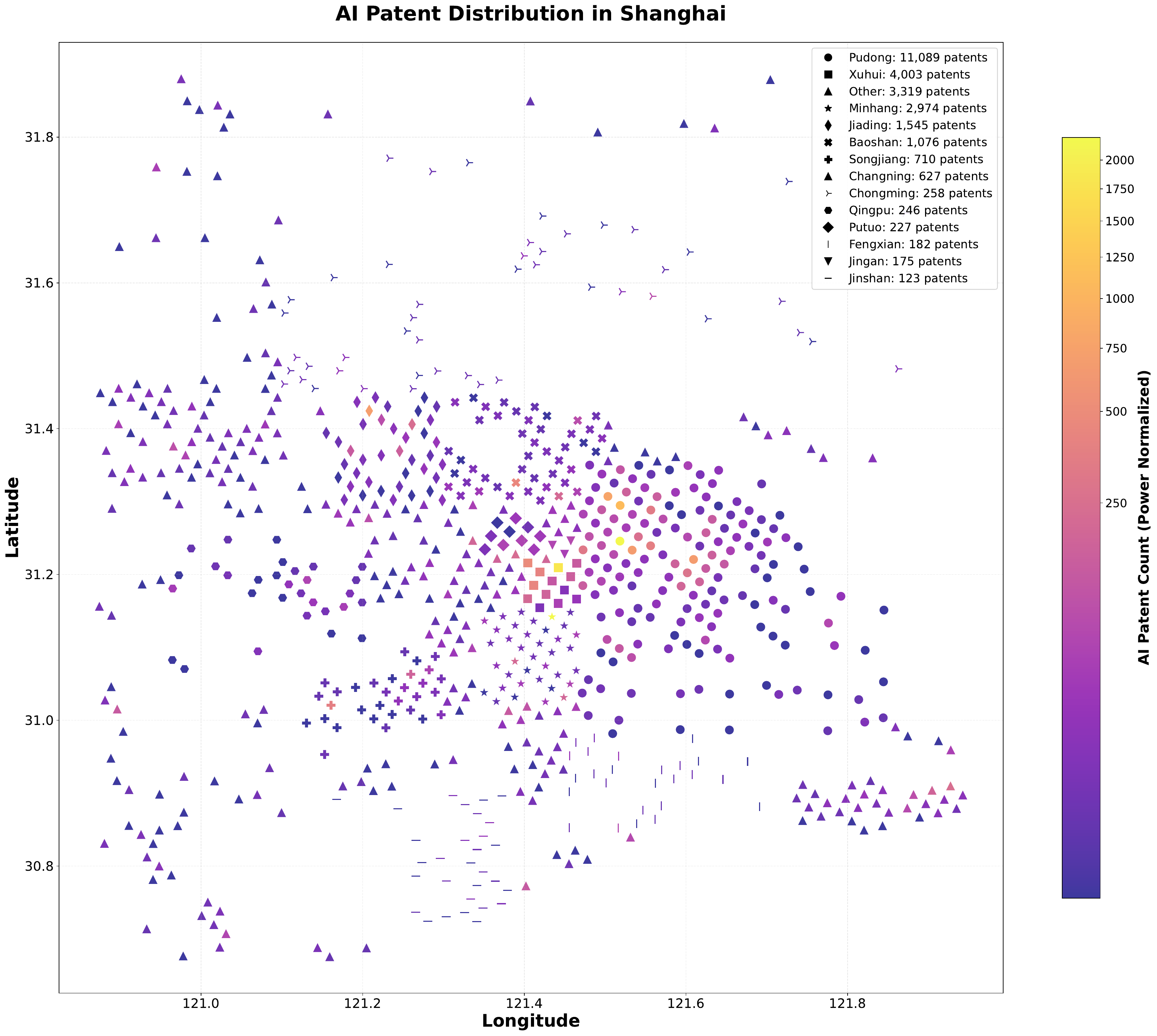}
        \caption{Panel B: Shanghai}
        \label{fig:panel_shanghai}
    \end{subfigure}

    \vspace{1.5em}

    \begin{subfigure}[t]{0.45\textwidth}
        \centering
        \includegraphics[width=\textwidth]{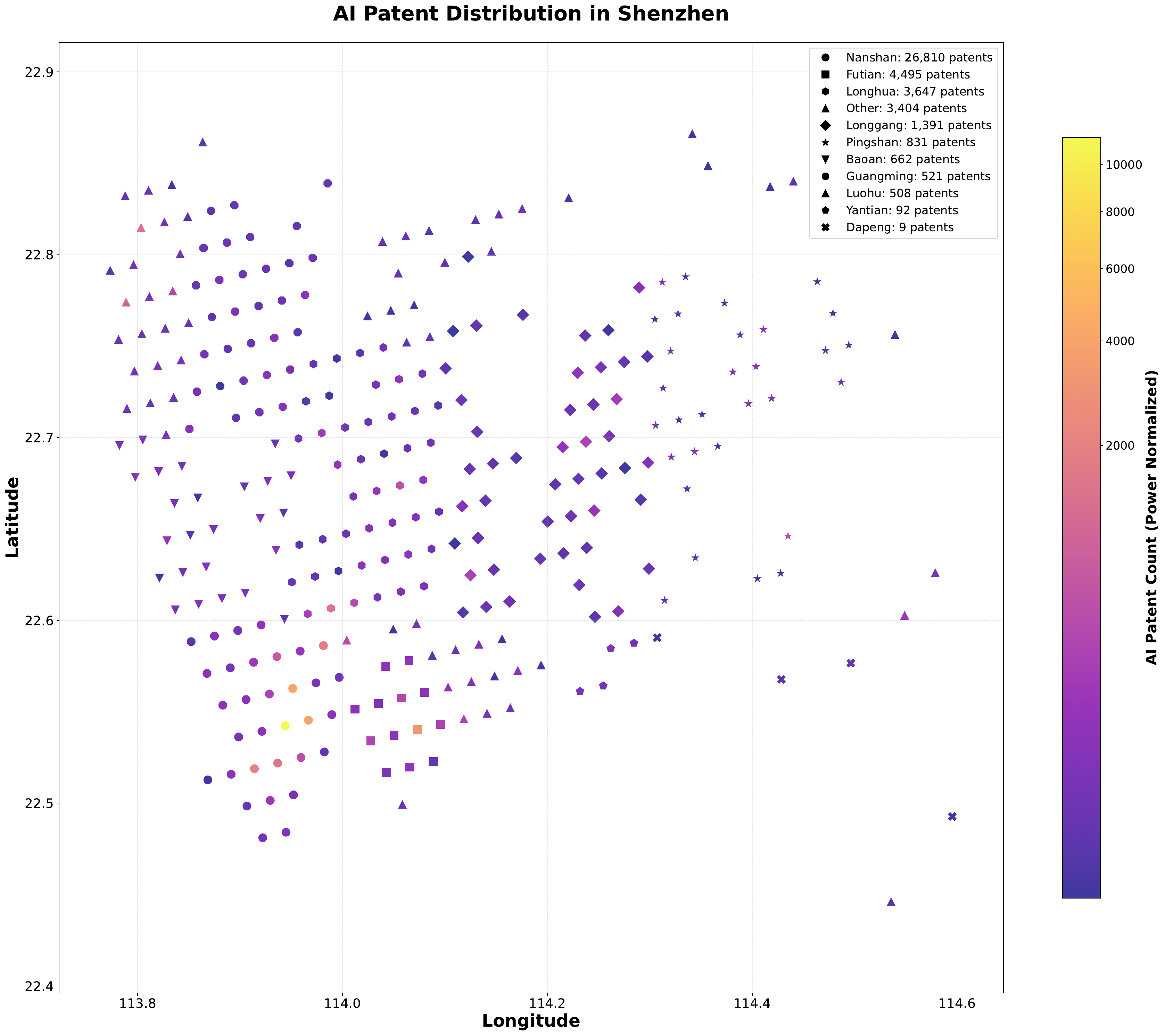}
        \caption{Panel C: Shenzhen}
        \label{fig:panel_shenzhen}
    \end{subfigure}

    \label{fig:ai_patent_city_maps}
\end{figure}
\restoregeometry

\renewcommand{\thetable}{B.\arabic{table}}
\setcounter{table}{0}

\newgeometry{left=1.2cm, right=1.2cm, top=2cm, bottom=2cm}
\begin{table}[htbp]
\centering
\caption{Restricting to AI-Active Grids}
\label{tab:a2}
\caption*{
This table reports estimates of changes in the number of new firms following the release of ChatGPT. For robustness, we restrict the sample to grids with at least one prior AI patent. \textit{Post} is an indicator equal to one for quarters after 2022Q4, and \textit{HighAI} is a dummy equal to one for grids with more than eight AI patents (the 75th-percentile threshold) filed during the period 2010–2019. Columns (1)–(3) use the following dependent variables: (1) number of new firms; (2) number of small firms; and (3) number of large firms. All regressions include city-by-quarter fixed effects and grid-by-calendar-quarter fixed effects to absorb seasonal patterns.  Standard errors are clustered at the city level. Variable definitions are provided in  Section~\ref{tab:variable_definition}. \textit{Notes:} *** $p<0.01$, ** $p<0.05$, * $p<0.1$.
}
\renewcommand{\arraystretch}{1} 
    \begin{tabular}{p{4cm}*{3}{>{\centering\arraybackslash}p{3cm}}}
        \toprule
        \input{appendix_tables/appendix_a2}
    \end{tabular}
\end{table}
\newpage
\restoregeometry

\newgeometry{left=1.2cm, right=1.2cm, top=2cm, bottom=2cm}
\begin{table}[htbp]
\centering
\caption{Matched Comparison with Nearby Non-AI Grids}
\label{tab:a4}
\caption*{
This table reports estimates of changes in the number of new firms following the release of ChatGPT. As a robustness check, we construct the sample by matching each grid with at least one AI patent to the five geographically closest grids without AI patents (with replacement). \textit{Post} is an indicator equal to one for quarters after 2022Q4, and \textit{HighAI} is a dummy equal to one for grids with at least one AI patent filed between 2010 and 2019. Columns (1)–(3) use the following dependent variables: (1) number of new firms; (2) number of small firms; and (3) number of large firms. All regressions include city-by-quarter fixed effects and grid-by-calendar-quarter fixed effects to absorb seasonal patterns. Standard errors are clustered at the city level. Variable definitions are provided in  Section~\ref{tab:variable_definition}. \textit{Notes:} *** $p<0.01$, ** $p<0.05$, * $p<0.1$.
}
\renewcommand{\arraystretch}{1} 
    \begin{tabular}{p{4cm}*{3}{>{\centering\arraybackslash}p{3cm}}}
        \toprule
        \input{appendix_tables/appendix_a4}
    \end{tabular}
\end{table}
\newpage
\restoregeometry

\newgeometry{left=1.2cm, right=1.2cm, top=2cm, bottom=2cm}
\begin{table}[htbp]
\centering
\caption{Changing Capital Threshold for Firm Size}
\label{tab:threshold}
\caption*{
This table reports estimates of changes in the number of new firms with different levels of registered capital following the release of ChatGPT. As a robustness check, we vary the cutoff used to define small firms. Specifically, in Columns (1)–(2), small firms are defined as firms with registered capital below 2 million RMB, with larger firms classified as large firms. In Columns (3)–(4), the cutoff is set at 3 million RMB, and in Columns (5)–(6), at 5 million RMB. \textit{Post} is an indicator equal to one for quarters after 2022Q4, and \textit{HighAI} is a dummy equal to one for grids with at least one AI patent. All specifications include city-by-quarter fixed effects and grid-by-calendar-quarter fixed effects. Variable definitions are provided in  Section~\ref{tab:variable_definition}. \textit{Notes:} *** $p<0.01$, ** $p<0.05$, * $p<0.1$.
}
\renewcommand{\arraystretch}{1} 
    \resizebox{\textwidth}{!}{
    \begin{tabular}{p{4cm}*{8}{>{\centering\arraybackslash}p{2cm}}}
        \toprule
        \input{appendix_tables/appendix_a3}
    \end{tabular}
    }
\end{table}
\newpage
\restoregeometry

\end{document}

%% file: main_tables/main_t1.tex
Variable                     & N         & Mean      & sd         & Min & Median    & Max  \\
\midrule
Num new firms                & 2,658,496 & 4.822  & 39.626 & 0   & 0      & 9,678 \\
Num small firms              & 2,658,496 & 2.082  & 20.536 & 0   & 0      & 9,672 \\
Num large firms              & 2,658,496 & 2.469  & 22.179 & 0   & 0      & 4,262 \\
Num high upstream            & 2,658,496 & 0.969  & 11.010 & 0   & 0      & 1,772 \\
Num low upstream             & 2,658,496 & 3.850  & 30.386 & 0   & 0      & 9,677 \\
Num high downstream          & 2,658,496 & 2.494  & 24.761 & 0   & 0      & 3,949 \\
Num low downstream           & 2,658,496 & 2.325  & 18.700 & 0   & 0      & 9,652 \\
Num high entrep              & 2,658,496 & 3.617  & 34.748 & 0   & 0      & 8,504 \\
Num low entrep               & 2,658,496 & 1.202  & 6.997  & 0   & 0      & 1,856 \\
Pct serial entrep            & 1,041,042 & 26.743 & 33.560 & 0   & 12.5   & 100   \\
Pct serial entrep small      & 1,041,042 & 16.346 & 30.013 & 0   & 0      & 100   \\
Pct serial entrep large      & 1,041,042 & 19.936 & 32.136 & 0   & 0      & 100   \\
Tot shareholders             & 1,002,793 & 1.502  & 0.852  & 1   & 1.333  & 139   \\
Tot shareholders small       & 670,018   & 1.389  & 0.693  & 1   & 1.103  & 49    \\
Tot shareholders large       & 743,846   & 1.624  & 0.991  & 1   & 1.5    & 139   \\
Pct indiv shareholders       & 1,002,793 & 93.205 & 18.650 & 0   & 100    & 100   \\
Pct indiv shareholders small & 670,018   & 96.057 & 14.851 & 0   & 100    & 100   \\
Pct indiv shareholders large & 743,846   & 90.705 & 21.807 & 0   & 100    & 100   \\
Exec team size               & 934,016   & 2.031  & 0.636  & 1   & 2      & 22    \\
Exec team size small         & 595,785   & 1.965  & 0.608  & 1   & 2      & 19    \\
Exec team size large         & 714,361   & 2.095  & 0.662  & 1   & 2      & 32 \\
AIpat                & 2,658,496 & 2.051  & 58.812 & 0   & 0      & 11,279 \\
HighAI                & 2,658,496 & 0.061  & 0.240 & 0   & 0      & 1 \\
\bottomrule

%% file: main_tables/main_t2.tex
               &\multicolumn{1}{c}{(1)}&\multicolumn{1}{c}{(2)}&\multicolumn{1}{c}{(3)} \\
               &\multicolumn{1}{c}{Num new firms} 
               & \multicolumn{1}{c}{Num small firms} & \multicolumn{1}{c}{Num large firms} \\              
\midrule
Post × HighAI  &       5.038\sym{***}&       
7.704\sym{***}&      -3.120\sym{***}\\
               &     (1.395)         &     
               (1.222)         &     (0.671)         \\
[1em]
Constant       &       4.668\sym{***}&       
1.847\sym{***}&       2.564\sym{***}\\
               &    (0.0427)         &    
               (0.0374)         &    (0.0205)         \\
\midrule
Observations   &     2,658,304         &    
2,658,304         &     2,658,304         \\
R-squared      &       0.810         &      
0.625         &       0.790         \\
City × Quarter FE&         Yes         &        
Yes         &         Yes         \\
Grid × Cal QTR FE&         Yes         &         
Yes         &         Yes         \\
\bottomrule

%% file: main_tables/main_t3.tex
               &\multicolumn{3}{c}{Winsorize by quarter}                                               &\multicolumn{3}{c}{Winsorize across all data}                                          \\\cmidrule(lr){2-4}\cmidrule(lr){5-7}
               &\multicolumn{1}{c}{(1)}&\multicolumn{1}{c}{(2)}&\multicolumn{1}{c}{(3)}&\multicolumn{1}{c}{(4)}&\multicolumn{1}{c}{(5)}&\multicolumn{1}{c}{(6)} \\
               &\multicolumn{1}{c}{Num new firms} 
               & \multicolumn{1}{c}{Num small firms} & \multicolumn{1}{c}{Num large firms} &\multicolumn{1}{c}{Num new firms} 
               & \multicolumn{1}{c}{Num small firms} & \multicolumn{1}{c}{Num large firms} \\    
\midrule
Post × HighAI  &       3.692\sym{***}&      
6.170\sym{***}&      -2.928\sym{***}&       2.518\sym{***}&    
4.911\sym{***}&      -2.922\sym{***}\\
               &     (0.826)         &     
               (0.800)         &     (0.332)         &     (0.667)         &     
               (0.528)         &     (0.336)         \\
[1em]
Constant       &       4.275\sym{***}&      
1.677\sym{***}&       2.321\sym{***}&       4.304\sym{***}&    
1.694\sym{***}&       2.307\sym{***}\\
               &    (0.0253)         &    
               (0.0245)         &    (0.0102)         &    (0.0204)         &    
               (0.0162)         &    (0.0103)         \\
\midrule
Observations   &     2,658,304         &     2,658,304         &     2,658,304         &     2,658,304         &     2,658,304         &     2,658,304                \\
R-squared      &       0.876         &      
0.771         &       0.856         &       0.885         &     
0.811         &       0.875         \\
City × Quarter FE&         Yes         &         Yes         &         Yes         &         Yes         &         Yes         &         Yes           \\
Grid × Cal QTR FE&         Yes         &         Yes         &         Yes         &         Yes         &         Yes         &         Yes         \\
\bottomrule

%% file: main_tables/main_t5.tex
Industry Name & Firm Count & Coefficient & SE & p-value \\
\midrule
Retail Industry & 2,019,644 & 1.6256 & 0.5447 & 0.0030 \\
Business Services & 1,135,654 & 0.9771 & 0.2921 & 0.0009 \\
Technology Promotion and Application Services & 1,301,263 & 0.8712 & 0.2316 & 0.0002 \\
Wholesale Industry & 2,061,046 & 0.4547 & 0.2466 & 0.0661 \\
Entertainment Industry & 137,133 & 0.2223 & 0.0534 & 0.0000 \\
Catering Industry & 199,082 & 0.1970 & 0.0389 & 0.0000 \\
Culture and Arts Industry & 289,160 & 0.1806 & 0.0579 & 0.0020 \\
Resident Services Industry & 169,229 & 0.1464 & 0.0294 & 0.0000 \\
Internet and Related Services & 118,450 & 0.0963 & 0.0277 & 0.0006 \\
Software and IT Services & 503,900 & 0.0556 & 0.1064 & 0.6016 \\
Broadcasting, Television, Film, and Audio Production & 48,885 & 0.0443 & 0.0155 & 0.0045 \\
Electricity, Heat, Gas, and Water Supply & 42,173 & 0.0400 & 0.0055 & 0.0000 \\
Leasing Industry & 152,152 & 0.0388 & 0.0215 & 0.0720 \\
Agriculture & 233,704 & 0.0360 & 0.0172 & 0.0370 \\
Chemical Raw Materials and Chemical Products Manufacturing & 28,633 & 0.0340 & 0.0149 & 0.0232 \\
\bottomrule

%% file: main_tables/main_t5_bottom.tex
Industry Name & Firm Count & Coefficient & SE & p-value \\
\midrule
Building Construction & 217{,}478 & -0.1588 & 0.1063 & 0.1363 \\
Civil Engineering Construction & 256{,}719 & -0.0858 & 0.0372 & 0.0218 \\
Decoration, Renovation, and Other Construction & 235{,}693 & -0.0565 & 0.0148 & 0.0002 \\
Construction Installation & 106{,}877 & -0.0194 & 0.0172 & 0.2591 \\
Real Estate Industry & 243{,}165 & -0.0136 & 0.0166 & 0.4108 \\
Other Manufacturing & 25{,}799 & -0.0096 & 0.0067 & 0.1492 \\
Education & 82{,}810 & -0.0089 & 0.0119 & 0.4569 \\
Metal Products Manufacturing & 67{,}641 & -0.0079 & 0.0038 & 0.0381 \\
General Equipment Manufacturing & 55{,}450 & -0.0066 & 0.0036 & 0.0665 \\
Computer, Communication, and Electronics Manufacturing & 23{,}854 & -0.0054 & 0.0095 & 0.5653 \\
Ecological and Environmental Governance & 17{,}471 & -0.0053 & 0.0021 & 0.0102 \\
Nonmetallic Mineral Products & 48{,}849 & -0.0041 & 0.0028 & 0.1460 \\
Monetary and Financial Services & 8{,}630 & -0.0032 & 0.0013 & 0.0169 \\
Specialized Equipment Manufacturing & 42{,}793 & -0.0029 & 0.0054 & 0.5898 \\
Instrumentation Manufacturing & 9{,}486 & -0.0020 & 0.0014 & 0.1536 \\
\bottomrule

%% file: main_tables/main_t4.tex
               &\multicolumn{1}{c}{(1)}&\multicolumn{1}{c}{(2)}&\multicolumn{1}{c}{(3)}&\multicolumn{1}{c}{(4)}&\multicolumn{1}{c}{(5)}&\multicolumn{1}{c}{(6)}\\
            &\multicolumn{1}{c}{Num high upstream}&\multicolumn{1}{c}{Num low upstream}&\multicolumn{1}{c}{Num high downstream}&\multicolumn{1}{c}{Num low downstream}&\multicolumn{1}{c}{Num high entrep}&\multicolumn{1}{c}{Num low entrep}\\
\midrule
Post × HighAI  &       0.846\sym{**} &       4.199\sym{***}&       3.323\sym{***}&       1.722\sym{**} &       5.892\sym{***}&      -0.848\sym{***}\\
               &     (0.388)         &     (1.225)         &     (0.789)         &     (0.753)         &     (1.302)         &     (0.300)         \\
[1em]
Constant       &       0.944\sym{***}&       3.721\sym{***}&       2.393\sym{***}&       2.272\sym{***}&       3.437\sym{***}&       1.228\sym{***}\\
               &    (0.0119)         &    (0.0375)         &    (0.0242)         &    (0.0231)         &    (0.0399)         &   (0.00918)         \\
\midrule
Observations   &     2,658,304         &     2,658,304         &     2,658,304         &     2,658,304         &     2,658,304         &     2,658,304         \\
R-squared      &       0.834         &       0.770         &       0.857         &       0.638         &       0.797         &       0.746         \\
City × Quarter FE&         Yes         &         Yes         &         Yes         &         Yes         &         Yes         &         Yes         \\
Grid × Cal QTR FE&         Yes         &         Yes         &         Yes         &         Yes         &         Yes         &         Yes         \\
\bottomrule

%% file: main_tables/main_t6.tex
               &\multicolumn{1}{c}{(1)}&\multicolumn{1}{c}{(2)}&\multicolumn{1}{c}{(3)}\\ 
               &\multicolumn{1}{c}{Pct serial entrep} & 
               \multicolumn{1}{c}{Pct serial entrep small} & \multicolumn{1}{c}{Pct serial entrep large} \\ 
\midrule
Post × HighAI  &      -0.405\sym{**} &      
-2.456\sym{***}&       0.648\sym{**} \\
               &     (0.201)         &     
               (0.266)         &     (0.294)         \\
[1em]
Constant       &       27.47\sym{***}&       
17.82\sym{***}&       21.25\sym{***}\\
               &    (0.0116)         &    
               (0.0154)         &    (0.0170)         \\
\midrule
Observations   &      863,869         &     
863,869         &      863,869         \\
R-squared      &       0.433         &      
0.420         &       0.425         \\
City × Quarter FE&         Yes         &       
Yes         &         Yes         \\
Grid × Cal QTR FE&         Yes         &         
Yes         &         Yes         \\
\bottomrule

%% file: main_tables/main_t7_panelA.tex
               &\multicolumn{1}{c}{(1)}&\multicolumn{1}{c}{(2)}&\multicolumn{1}{c}{(3)}\\ 
               &\multicolumn{1}{c}{Tot shareholders} & 
               \multicolumn{1}{c}{Tot shareholders small} & \multicolumn{1}{c}{Tot shareholders large} \\  
\midrule
Post × HighAI  &     -0.0211\sym{***}&     
-0.0134\sym{*}  &     -0.0173\sym{**} \\
               &   (0.00747)         &   
               (0.00755)         &   (0.00781)         \\
[1em]
Constant       &       1.513\sym{***}&      
1.412\sym{***}&       1.639\sym{***}\\
               &  (0.000444)         &  
               (0.000646)         &  (0.000598)         \\
\midrule
Observations   &      827,286         &     
512,172         &      585,030         \\
R-squared      &       0.423         &      
0.434         &       0.427         \\
City × Quarter FE&         Yes         &       
Yes         &         Yes         \\
Grid × Cal QTR FE&         Yes         &        
Yes         &         Yes         \\
\bottomrule

%% file: main_tables/main_t7_panelB.tex
               &\multicolumn{1}{c}{(1)}&\multicolumn{1}{c}{(2)}&\multicolumn{1}{c}{(3)}\\ 
               &\multicolumn{1}{c}{Pct indiv shareholders} & 
               \multicolumn{1}{c}{\makecell{Pct indiv \\ shareholders small}} & \multicolumn{1}{c}{\makecell{Pct indiv \\ shareholders large}} \\  
\midrule
Post × HighAI  &      0.0578         &      
-0.185         &     -0.0774         \\
               &     (0.161)         &     
               (0.132)         &     (0.169)         \\
[1em]
Constant       &       93.29\sym{***}&      
96.22\sym{***}&       90.55\sym{***}\\
               &   (0.00955)         &   
               (0.0113)         &    (0.0130)         \\
\midrule
Observations   &      827,286         &     
512,172         &      585,030         \\
R-squared      &       0.456         &      
0.434         &       0.472         \\
City × Quarter FE&         Yes         &      
Yes         &         Yes         \\
Grid × Cal QTR FE&         Yes         &        
Yes         &         Yes         \\
\bottomrule

%% file: main_tables/main_t8.tex
               &\multicolumn{1}{c}{(1)}&\multicolumn{1}{c}{(2)}&\multicolumn{1}{c}{(3)}\\ 
               &\multicolumn{1}{c}{Exec team size} & 
               \multicolumn{1}{c}{Exec team size small} & \multicolumn{1}{c}{Exec team size large} \\  
\midrule
Post × HighAI  &     -0.0161\sym{***}&     
-0.0187\sym{***}&    -0.00373         \\
               &   (0.00392)         &   
               (0.00401)         &   (0.00441)         \\
[1em]
Constant       &       2.032\sym{***}&       
1.976\sym{***}&       2.091\sym{***}\\
               &  (0.000249)         &  
               (0.000382)         &  (0.000349)         \\
\midrule
Observations   &      765,404         &     
452,023         &      560,617         \\
R-squared      &       0.568         &      
0.611         &       0.546         \\
City × Quarter FE&         Yes         &        
Yes         &         Yes         \\
Grid × Cal QTR FE&         Yes         &        
Yes         &         Yes         \\
\bottomrule

%% file: appendix_tables/appendix_a5_panelA.tex
               &\multicolumn{1}{c}{(1)}&\multicolumn{1}{c}{(2)}&\multicolumn{1}{c}{(3)}\\
               &\multicolumn{1}{c}{All firms}&\multicolumn{1}{c}{Small firms}&\multicolumn{1}{c}{Large firms}\\
\midrule
Post × HighAI  &      -0.629\sym{*}  &      -5.416\sym{***}&      -0.917\sym{***}\\
               &     (0.320)         &     (0.483)         &     (0.241)         \\
[1em]
Constant       &       36.29\sym{***}&       40.95\sym{***}&       17.73\sym{***}\\
               &    (0.0348)         &    (0.0525)         &    (0.0262)         \\
\midrule
Observations   &      375,548         &      375,548         &      375,548         \\
R-squared      &       0.411         &       0.438         &       0.422         \\
City × Quarter FE&         Yes         &         Yes         &         Yes         \\
Grid × Cal QTR FE&         Yes         &         Yes         &         Yes         \\
\bottomrule

%% file: appendix_tables/appendix_a5_panelB.tex
               &\multicolumn{1}{c}{(1)}&\multicolumn{1}{c}{(2)}&\multicolumn{1}{c}{(3)}\\
               &\multicolumn{1}{c}{All firms}&\multicolumn{1}{c}{Small firms}&\multicolumn{1}{c}{Large firms}\\
\midrule
Post × HighAI  &      -1.983\sym{***}&      -7.100\sym{***}&      -0.180\sym{***}\\
               &     (0.560)         &     (1.148)         &    (0.0230)         \\
[1em]
Constant       &       9.601\sym{***}&       16.64\sym{***}&       1.200\sym{***}\\
               &    (0.0612)         &     (0.124)         &   (0.00249)         \\
\midrule
Observations   &      366,604         &      370,610         &      370,861         \\
R-squared      &       0.375         &       0.359         &       0.405         \\
City × Quarter FE&         Yes         &         Yes         &         Yes         \\
Grid × Cal QTR FE&         Yes         &         Yes         &         Yes         \\
\bottomrule

%% file: main_tables/main_t10.tex
               &\multicolumn{1}{c}{(1)}&\multicolumn{1}{c}{(2)}&\multicolumn{1}{c}{(3)}\\ 
               &\multicolumn{1}{c}{Num new firms} 
               & \multicolumn{1}{c}{Num small firms} & \multicolumn{1}{c}{Num large firms} \\  
\midrule
Post × High nonAI&       0.159         &       0.558         &      -0.427\sym{*}  \\
               &     (0.478)         &     (0.485)         &     (0.249)         \\
[1em]
Constant       &       16.69\sym{***}&       7.063\sym{***}&       8.721\sym{***}\\
               &    (0.0589)         &    (0.0599)         &    (0.0308)         \\
\midrule
Observations   &      546,768         &      546,768         &      546,768          \\
R-squared      &       0.818         &       0.645         &       0.791         \\
City × Quarter FE&         Yes         &         Yes         &         Yes         \\
Grid × Cal QTR FE&         Yes         &         Yes         &         Yes         \\
\bottomrule

%% file: main_tables/main_t11.tex
               &\multicolumn{1}{c}{(1)}&\multicolumn{1}{c}{(2)}&\multicolumn{1}{c}{(3)}\\
               &\multicolumn{1}{c}{Num new firms}&\multicolumn{1}{c}{Num small firms}&\multicolumn{1}{c}{Num large firms}\\
\midrule
Post × HighResid&       0.782\sym{***}&       1.216\sym{***}&      -0.481\sym{***}\\
               &     (0.173)         &     (0.143)         &    (0.0601)         \\
[1em]
Constant       &       4.629\sym{***}&       1.782\sym{***}&       2.587\sym{***}\\
               &    (0.0427)         &    (0.0352)         &    (0.0149)         \\
\hline
Observations   &     2,658,304         &     2,658,304         &     2,658,304         \\
R-squared      &       0.810         &       0.623         &       0.790         \\
City × Quarter FE&         Yes         &         Yes         &         Yes         \\
Grid × Cal QTR FE&         Yes         &         Yes         &         Yes         \\
\bottomrule

%% file: main_tables/main_t9.tex
               &\multicolumn{1}{c}{(1)}&\multicolumn{1}{c}{(2)}&\multicolumn{1}{c}{(3)}\\ 
               &\multicolumn{1}{c}{Num new firms} 
               & \multicolumn{1}{c}{Num small firms} & \multicolumn{1}{c}{Num large firms} \\  
\midrule
Post × HighAI  &       3.979\sym{**} &      
6.288\sym{***}&      -2.684\sym{***}\\
               &     (1.537)         &     
               (1.165)         &     (0.761)         \\
[1em]
Constant       &       3.512\sym{***}&       
1.352\sym{***}&       1.942\sym{***}\\
               &    (0.0383)         &    
               (0.0291)         &    (0.0190)         \\
\midrule
Observations   &     2,477,248         &     2,477,248         &     2,477,248             \\
R-squared      &       0.758         &      
0.510         &       0.802         \\
City × Quarter FE&         Yes         &         Yes         &         Yes         \\
Grid × Cal QTR FE&         Yes          &         Yes         &         Yes         \\
\bottomrule

%% file: main_tables/variable_def.tex
\textbf{Variable} & \textbf{Definition} \\
\midrule
Num new firms                & Number of new firms in grid $g$ during quarter $t$.                                                                                                                                                                                           \\
Num small firms              & Number of small firms (registered capital $<$ 1 million RMB) in grid $g$ during quarter $t$.                                                                                                                                             \\
Num large firms              & Number of large firms (registered capital $\ge$ 1 million RMB) in grid $g$ during quarter $t$.                                                                                                                                           \\
Num high upstream            & Number of new firms with AI upstream score $>$ 0 in grid $g$ during quarter $t$.                                                                                                                                                              \\
Num low upstream             & Number of new firms with AI upstream score $\leq$ 0 in grid $g$ during quarter $t$.                                                                                                                                                           \\
Num high downstream          & Number of new firms with AI downstream score $>$ 0 in grid $g$ during quarter $t$.                                                                                                                                                            \\
Num low downstream           & Number of new firms with AI downstream score $\leq$ 0 in grid $g$ during quarter $t$.                                                                                                                                                         \\
Num high entrep              & Number of new firms with Entrepreneurship Helpfulness score $>$ 0 in grid $g$ during quarter $t$.                                                                                                                                             \\
Num low entrep               & Number of new firms with Entrepreneurship Helpfulness score $\leq$ 0 in grid $g$ during quarter $t$.                                                                                                                                          \\
Pct serial entrep            & Average percentage of serial entrepreneurs (legal representatives who had founded at least one other firm within the three years preceding the current firm’s establishment, excluding the month of founding) in grid $g$ during quarter $t$. \\
Pct serial entrep small      & Average percentage of serial entrepreneurs in small businesses in grid $g$ during quarter $t$.                                                                                                                                                \\
Pct serial entrep large      & Average percentage of serial entrepreneurs in large businesses in grid $g$ during quarter $t$.                                                                                                                                                \\
Tot shareholders             & Average number of shareholders of new firms in grid $g$ during quarter $t$.                                                                                                                                                                   \\
Tot shareholders small       & Average number of shareholders of small businesses in grid $g$ during quarter $t$.                                                                                                                                                            \\
Tot shareholders large       & Average number of shareholders of large businesses in grid $g$ during quarter $t$.                                                                                                                                                            \\
Pct indiv shareholders       & Average number of individual shareholders divided by total shareholders × 100 for new firms in grid $g$ during quarter $t$.                                                                                                                   \\
Pct indiv shareholders small & Average number of individual shareholders divided by total shareholders × 100 for small businesses in grid $g$ during quarter $t$.                                                                                                            \\
Pct indiv shareholders large & Average number of individual shareholders divided by total shareholders × 100 for large businesses in grid $g$ during quarter $t$.                                                                                                            \\
Exec team size               & Average number of executive members of new firms in grid $g$ during quarter $t$.                                                                                                                                                              \\
Exec team size small         & Average number of executive members of small businesses in grid $g$ during quarter $t$.                                                                                                                                                       \\
Exec team size large         & Average number of executive members of large businesses in grid $g$ during quarter $t$. \\
AIpat & Number of AI-related patents
filed between 2010 and 2019 in grid $g$ \\
HighAI & Binary indicator equal to 1 if grid $g$ has at least one AI-related patent \\
\bottomrule

%% file: appendix_tables/appendix_a2.tex
               &\multicolumn{1}{c}{(1)}&\multicolumn{1}{c}{(2)}&\multicolumn{1}{c}{(3)}\\ 
               &\multicolumn{1}{c}{Num new firms}
               & \multicolumn{1}{c}{Num small firms} & \multicolumn{1}{c}{Num large firms} \\  
\midrule
Post × HighAI  &       6.568\sym{***}&      
11.62\sym{***}&      -5.859\sym{***}\\
               &     (1.962)         &     
               (2.815)         &     (1.942)         \\
[1em]
Constant       &       40.81\sym{***}&      
16.04\sym{***}&       22.69\sym{***}\\
               &     (0.235)         &     
               (0.338)         &     (0.233)         \\
\midrule
Observations   &      162,512         &      162,512         &      162,512       \\
R-squared      &       0.842         &      
0.709         &       0.796         \\
City × Quarter FE&         Yes         &         Yes         &         Yes          \\
Grid × Cal QTR FE&         Yes         &         Yes         &         Yes         \\
\bottomrule

%% file: appendix_tables/appendix_a4.tex
               &\multicolumn{1}{c}{(1)}&\multicolumn{1}{c}{(2)}&\multicolumn{1}{c}{(3)}\\ 
               &\multicolumn{1}{c}{Num new firms} 
               & \multicolumn{1}{c}{Num small firms} & \multicolumn{1}{c}{Num large firms} \\  
\midrule
Post × HighAI  &       4.345\sym{***}&      
6.584\sym{***}&      -2.621\sym{***}\\
               &     (1.354)         &     
               (1.484)         &     (0.693)         \\
[1em]
Constant       &       18.17\sym{***}&      
7.333\sym{***}&       9.889\sym{***}\\
               &     (0.113)         &     
               (0.124)         &    (0.0578)         \\
\midrule
Observations   &      977,408         &      977,408         &      977,408          \\
R-squared      &       0.794         &      
0.615         &       0.772         \\
City × Quarter FE&         Yes         &         Yes         &         Yes            \\
Grid × Cal QTR FE&         Yes         &         Yes         &         Yes         \\
\bottomrule

%% file: appendix_tables/appendix_a3.tex
               &\multicolumn{2}{c}{Cutoff 2 Millions}      &\multicolumn{2}{c}{Cutoff 3 Millions}      &\multicolumn{2}{c}{Cutoff 5 Millions}      \\\cmidrule(lr){2-3}\cmidrule(lr){4-5}\cmidrule(lr){6-7}
               &\multicolumn{1}{c}{(1)}&\multicolumn{1}{c}{(2)}&\multicolumn{1}{c}{(3)}&\multicolumn{1}{c}{(4)}&\multicolumn{1}{c}{(5)}&\multicolumn{1}{c}{(6)}\\
               & \multicolumn{1}{c}{Num small firms} & \multicolumn{1}{c}{Num large firms} & \multicolumn{1}{c}{Num small firms} & \multicolumn{1}{c}{Num large firms} & \multicolumn{1}{c}{Num small firms} & \multicolumn{1}{c}{Num large firms} \\
\midrule
Post × HighAI  &       7.680\sym{***}&      -3.096\sym{***}&       7.380\sym{***}&      -2.796\sym{***}&       7.157\sym{***}&      -2.573\sym{***}\\
               &     (1.368)         &     (0.473)         &     (1.384)         &     (0.440)         &     (1.386)         &     (0.421)         \\
[1em]
Constant       &       2.977\sym{***}&       1.434\sym{***}&       3.319\sym{***}&       1.092\sym{***}&       3.536\sym{***}&       0.875\sym{***}\\
               &    (0.0419)         &    (0.0145)         &    (0.0424)         &    (0.0135)         &    (0.0425)         &    (0.0129)         \\
\midrule
Observations   &     2,658,304         &     2,658,304         &     2,658,304         &     2,658,304         &     2,658,304         &     2,658,304         \\
R-squared      &       0.752         &       0.756         &       0.762         &       0.751         &       0.769         &       0.731         \\
City × Quarter FE&         Yes         &         Yes         &         Yes         &         Yes         &         Yes         &         Yes         \\
Grid × Cal QTR FE&         Yes         &         Yes         &         Yes         &         Yes         &         Yes         &         Yes         \\
\bottomrule